\documentclass[aps,twocolumn,superscriptaddress]{revtex4-2}

\usepackage{bm}
\usepackage{times}
\usepackage{dsfont}
\usepackage{amssymb}
\usepackage{amsmath,amsthm,amsbsy}
\usepackage{graphicx}
\usepackage{xcolor}
\usepackage{ bbold }
\usepackage{upgreek}

\usepackage[colorlinks,linkcolor=blue,citecolor=blue,anchorcolor=blue,urlcolor=blue]{hyperref}

\newcommand{\bl}{\begin{aligned}}
\newcommand{\el}{\end{aligned}}

\def\be{\begin{equation}}
\def\ee{\end{equation}}

\def\bi{\begin{itemize}}
\def\ei{\end{itemize}}
\def\bn{\begin{enumerate}}
\def\en{\end{enumerate}}
\def\bea{\begin{eqnarray}}
\def\eea{\end{eqnarray}}
\def\no{\nonumber}
\def\ba{\begin{array}}
\def\ea{\end{array}}
\def\bd{\begin{displaymath}}
\def\ed{\end{displaymath}}

\graphicspath{{Figs/}}

\begin{document}

\title{Dynamics of quantum Fisher 
and Wigner-Yanase skew information following a noisy quench}

\author{J. Naji}
\email[]{j.naji@ilam.ac.ir}
\affiliation{Department of Physics, Faculty of Science, Ilam University, Ilam, Iran}
\author{R. Jafari}
\email{raadmehr.jafari@gmail.com}
\affiliation{Physics Department and Research Center OPTIMAS, University of Kaiserslautern, 67663 Kaiserslautern, Germany}
\author{Alireza  Akbari}
\email{alireza@bimsa.cn}
\affiliation{Beijing Institute of Mathematical Sciences and Applications (BIMSA), Huairou District, Beijing 101408, China}

\author{M. Abdi}
\email[]{mehabdi@gmail.com}
\affiliation{School of Physics and Astronomy, Shanghai Jiao Tong University, Shanghai 200240, China}

\date{\today}

\begin{abstract}
We study the effect of noise on the dynamics of the transverse-field Ising model quenched across a quantum critical point. 
To quantify two-spin correlations, we employ the quantum Fisher information (QFI) and the Wigner--Yanase skew information (WYSI) as measures of quantum coherence. 
In the noiseless case, in contrast to the dynamics of entanglement in anisotropic XY chains, both QFI and WYSI increase monotonically with the ramp quench time, approaching their adiabatic limits without exhibiting any Kibble–Zurek–type scaling with quench duration. 
In contrast, when  noise is added to the quench protocol, the coherence dynamics change qualitatively: QFI and WYSI both decay exponentially with the time scale of a ramp quench, with an exponent determined by the noise intensity. 
Furthermore, the maximum ramp time, at which either of these measures reach their maximum, scales linearly with the noise variance, featuring the same exponent that determines the optimal annealing time for minimizing defect production in noisy quantum annealing.  
\end{abstract}

\maketitle

\section{Introduction}
Quantum phase transitions (QPTs) have been one of the most important topics in the area of strongly correlated systems \cite{sachdevbook,Vojta2003,chakrabartibook}. 
There have been recent developments in theoretical investigation of QPTs from the viewpoint of quantum information theory. 
Quantities like concurrence \cite{Hill1997,osterloh2002}, entanglement entropy \cite{Vidal2003,Kitaev2006}, fidelity susceptibility 
\cite{Campos2007,Zanardi2007,Gu2008,Schwandt2009}, and geometric phases \cite{Carollo2005,Zhu2006} have been shown to capture the non-analyticities associated with a quantum critical point (QCP). 
Recent advances in the studies of ultracold atoms trapped in optical lattices have opened a new avenue of investigation of the non-equilibrium dynamics of QPTs in quantum many body systems \cite{greiner2002,Jaksch1998}.
Divergences in correlation length and time scales at QCPs introduce non-trivial effects in the dynamics of a quantum system near a QCP.
In particular, if a parameter of the Hamiltonian is tuned slowly so as to drive the system through a QCP, the dynamics fails to be adiabatic for any finite rate of variation of the parameter.
This non-adiabaticity in the response of the system eventually leads to generation of defects in the final state.
A quantitative analysis of defect production in a system following a quench can be carried out using the Kibble-Zurek (KZ) theory \cite{Kibble1976,ruutu1996} extended to quantum spin chains \cite{Zurek2005,Damski2005,Dziarmaga2005,Sen2008,Sengupta2008,Barankov2008,Subires2022,Bando2020,Bermudez2009}.

The recent studies provide a bridge between the nonequilibrium dynamics of a quantum critical system and quantum information theory 
\cite{Bertini2022,Sadiek2010,Mishra2018,Huang2006,Jafari2020a,Utkarsh2016,Ansari2024,Canovi2014,Miao2024,Paul2024,Sacramento2014,Pollmann2010,Su2020,Vincenzo2017,Bacsi2021,Eisler2007, Nag2011,Vincenzo2018,Romain2017,Pappalardi2017}. 
One may, for example, raise the question what is the value of quantum correlations in the final state of a quantum system following a quench across a QCP.
If the dynamics is perfectly adiabatic, then no additional correlation is generated in the final state. However, break down of adiabaticity in the passage through a second-order QCP is inevitable, and thus, leads to defects in the final state and these defects in turn lead to nonzero entanglement \cite{Cherng2006, Sengupta2009, Cincio2007}.
It has been shown that, the two-spin entanglement can be generated only for spins separated by even lattice spacing, for a very large field quench across two QCPs and the nearest-neighbor sites are not entangled \cite{Sengupta2009,Cherng2006}. 
In addition, it was established that two-spin entanglement also follows the same Kibble–Zurek scaling relation as the defect formation~\cite{Cherng2006, Cincio2007, Sengupta2009,Patra2011,Nag2011}.
However, comparatively little attention has been devoted to the stochastic driving of thermally isolated systems with noisy Hamiltonian and specifically, the role of quantum coherence remains largely unexplored. In any real experiment, the simulation of the desired time dependent Hamiltonian is imperfect and noisy fluctuations are inevitable.
In other words, the noises are ubiquitous and indispensable in any physical system e.g., the noise-induced heating, which can originate from amplitude fluctuations of the lasers forming the optical lattice~\cite{Zoller1981,Chen2010,Doria2011}.
Furthermore, noise is often used as a model for the effect of the environment, for instance a heat bath on a relatively small system. Understanding the effects of noise in such systems is of utmost importance both in designing experiments and comprehend the results \cite{Marino2012, Marino2014}.

This raises a fundamental question about the generation of quantum correlations after a noisy quench across a QCP. 
If such correlations are generated, how does noise influence their dynamical behavior? 
To address this, we analyze the dynamics of the quantum Fisher information (QFI) and the Wigner--Yanase skew information (WYSI) in the transverse-field Ising model (TFI) under both noiseless and noisy driving.
%
QFI \cite{Braunstein1994} is an extension of Fisher information \cite{Fisher1925} to the quantum realm, and is a fundamental concept of quantum metrology \cite{Roy2008,Boixo2007,Boixo2008,Giovannetti2006,Zhang2023}. The QFI plays an important role in quantum sensing and parameter estimation because it can provide a bound about the accuracy of quantum estimation, i.e., a larger QFI means a higher precision.
Apart from quantum metrology, the QFI also connects to other aspects of quantum physics, such as quantum phase transition \cite{Marzolino2017,Wang2014,Wang2017,Yin2019}, entanglement witness \cite{Hyllus2012,Geza2012,Li2013,Boixo2008,Pezze2009,Pezze2016}, uncertainty relations \cite{Gibilisco2007}, the calculation of quantum speedup limit time \cite{Taddei2013}.
Another quantum version of Fisher information is the WYSI~\cite{Wigner1963}, which has attracted considerable attention and is used to investigate entanglement~\cite{Chen2005,Li2006}.

In this work, we show that, in the noiseless case, both the two‐spin QFI and WYSI grow monotonically with the ramp time scale $\tau$, approaching their static (adiabatic) values for $\tau\to\infty$, and display characteristic extrema at the quantum critical fields $h_c=\pm1$ and at $h_0=0$ (for the longitudinal coherence).  
In contrast, when a Gaussian white noise of strength $\xi$ is added to the control field, the dynamics of both measures becomes nonmonotonic: each develops a clear maximum at an intermediate ramp time $\tau_m(\xi)$ beyond which coherence is suppressed by dephasing.  
By examining the logarithm of QFI and WYSI versus $\tau$, we establish that in the presence of noise both quantities decay exponentially with $\tau$, and that the $\tau$ which maximizes the information, $\tau_m(\xi)$, scales linearly with the noise variance $\xi^2$.  
These results uncover a direct quantitative link between noise strength and the dynamical generation and degradation of quantum correlations in critical many‐body systems. The identical scaling of QFI and WYSI with $\xi^2$ points to a common mechanism governing the competition between coherent defect production and noise‐induced dephasing. Such insights are of potential relevance for optimizing ramp protocols in quantum annealing and metrological schemes, where tuning the quench rate in the presence of unavoidable noise can maximize the usable quantum coherence.


\section{TIME DEPENDENT XY MODEL}
%
We consider the one‐dimensional spin-$1/2$ XY model of length $N$ in a transverse magnetic field $h(t)$, described by the Hamiltonian  
%
\be
{\cal H}(t){=}-\sum_{i=1}^{N} \Big( \frac{1{+}\gamma}{4}\sigma_{i}^{x} \sigma_{i+1}^{x} + \frac{1{-}\gamma}{4}\sigma_{i}^{y} \sigma_{i+1}^{y} +h(t)\sigma_i^z \Big),
\;
\label{eq1}
\ee
%
where $\sigma^{\alpha=\{x,y,z\}}_{i}$ are the Pauli matrices acting on site $i$, and $\gamma$ is the anisotropy parameter.
We impose periodic boundary conditions and, for numerical simulations, set \(N=200\).
In the case of a time-independent magnetic field ($h(t)=h$) the XY model exhibits an 
Ising‐type quantum phase transition at $\lvert h\rvert=1$,  and an anisotropic QPT along the line $\gamma=0$.
The Ising-like QPT occurs at $h=\pm1$ where the gap vanishes for $k=0$ and $k=\pi$ at the boundary between the paramagnetic phase and ferromagnetic phase.
By applying a Jordan–Wigner transformation \cite{LIEB,Jafari2012} followed by a Fourier transformation, the Hamiltonian in Eq.~\eqref{eq1} can be written as the sum of $N/2$ non-interacting terms
${\cal H}(t) = \sum_{k} {\cal H}_{k}(t)$,
%
%
with
%
\be
\bl
\label{eq:APA3}
{\cal H}_{k}(t)
= &
\Big(h(t)-\cos(k)\Big)\Big(c_{k}^{\dagger} c^{}_{k}+c_{-k}^{\dagger} c^{}_{-k}\Big)\\
&-
{\it i}\gamma\sin(k)\Big(c_{k}^{\dagger} c_{-k}^{\dagger}+c^{}_{k} c^{}_{-k}\Big),
\el
\ee
%
where $c_{k}^{\dagger}$ and  $c^{}_{k}$ are the spinless fermion creation and annihilation operators, respectively with $k= (2m-1)\pi/N$, for $m=1, 2, \dots, N/2$. 
Introducing the Nambu spinor $\mathds{C}^{\dagger}=(c_{k}^{\dagger},~c_{-k})$, the Hamiltonian ${\cal H}_{k}(t)$ can be expressed 
in Bogoliubov-de Gennes (BdG) form as
%
\be
\bl
\label{eq:APA4}
H_{k}(t)=\mathds{C}^{\dagger}{\cal H}_{k}(t)\mathds{C};
\quad
H_{k}(t)=\left(
\begin{array}{cc}
h_k(t) & -{\it i}\Delta_k \\
{\it i}\Delta_k & -h_k(t)\\
\end{array}
\right),
\el
\ee
%
where $h_k(t)=h(t)-\cos(k)$ and $\Delta_k=\gamma\sin(k)$. Therefore the Bloch single particle Hamiltonian can be represented as
$H_{k}(t)=h_k(t)\sigma^{z}+\Delta_k\sigma^{y}$ with the corresponding eigenenergies $\varepsilon_k^{\pm}(t)=\pm\varepsilon_{k}(t)=\pm\sqrt{h^2_k(t)+\Delta^2_k}$.
In the fermion excitation formalism the instantaneous eigenvalues and eigenvectors of Hamiltonian Eq. (\ref{eq:APA4}) are expressed as
%
\begin{subequations}
\label{eq:APA5}
\begin{align}
|\phi^{-}_{k}(t)\rangle&=\cos(\frac{\theta_k(t)}{2})|0\rangle-i\sin(\frac{\theta_k(t)}{2})c_{k}^{\dagger}c_{-k}^{\dagger}|0\rangle,\\
|\phi^{+}_{k}(t)\rangle&=-i\sin(\frac{\theta_k(t)}{2})|0\rangle+\cos(\frac{\theta_k(t)}{2})c_{k}^{\dagger}c_{-k}^{\dagger}|0\rangle,
\end{align}
\end{subequations}
%
where we have introduced $\theta_k(t)=\arctan[\Delta_k/h_k(t)]$.
\\

To study the quench dynamics, we employ a linear ramp protocol in the absence of noise by setting the time‐dependent field to $h(t)=h_0(t)=t/\tau$, where $\tau$ is the ramp duration. To include noise, we superimpose a stochastic term, writing $h(t)=h_0(t)+R(t)=t/\tau+R(t)$, where $R(t)$ is a zero‐mean random fluctuation ($\langle R({t})\rangle=0$) confined to the interval $[t_i,t_f]$.
Therefore, the Hamiltonian of each mode can be decomposed into $H_k(t)=H_{0,k}(t)+R(t)H_{1,k}$ with $H_{1,k}=\sigma^z$.
Here, we assume a white noise with Gaussian two-point correlations $\langle R(t)R(t')\rangle=\xi^2 \delta (t-t')$,
 where $\xi$ characterizes the noise strength.
Note that $\xi^2$ has units of time.
White noise is approximately equivalent to the fast colored noise with exponentially decaying two-point correlations (Ornstein-Uhlenbeck process)~\cite{Jafari2024}.
Here, we assume the system initially in a large magnetic field background, namely $h_i \ll h_c=-1$, where the ground state of the system is paramagnetic.
For this state all modes are initially in their lowest energy level, i.e., $|\psi(t=0)\rangle=\prod_{k>0}|\phi^{-}_k(t=0)\rangle$. 
Since the condition for an adiabatic dynamics breaks in the vicinity of the QCPs, after a ramp across the critical points $h_c=\pm1$ to any final value $h(t)>h_c$, the final state $|\psi_{k}(t) \rangle$ does not 
remain in the groundstate of the final Hamiltonian. Therefore, for the noiseless case the final state reads
%
\begin{equation} 
\label{eq:wavevector}
|\psi(t)\rangle=\prod_{k>0}|\psi_k(t)\rangle,
\end{equation}
%
where $|\psi_{k}(t) \rangle = v_k |\phi^{-}_k(t)\rangle+ u_k |\phi^{+}_k(t)\rangle$ with $|u_k|^2 + |v_k|^2 =1$, and $|u_k|^2$ is the probability of finding the $k$th mode in the excited state with energy $+\varepsilon_k$ at the end of the quench at time $t$.
Hence any correlations among the system components that are obtained under such a ramp is a nonadiabatic quench process and must therefore include contributions from the excited states of the system.
In the noiseless case, the amplitudes $u_k(t)$ and $v_k(t)$ are easily obtainable from the von Neumann equation 
\be
\dot{\rho}_{0,k}(t)=
{d\over dt}\rho_{0,k}(t)=
-i[H_{0,k}(t),\rho_{0,k}(t)].
\ee
Here, $\rho_{0,k}(t)$ is the density matrix of mode $k$ for the noiseless ramped quench, i.e, $h_k(t)=h_0(t)-\cos(k)$.
We need to solve this master equation to obtain the amplitudes $u_k(t)$ and $v_k(t)$ from the relations  
%
\be
\bl
\label{eq:amplitudes}
|u_k(t)|^2
&=
\langle \phi^{+}_{k}(t)|\rho_{0,k}(t)|\phi^{+}_{k}(t)\rangle, 
\\
v_{k}(t)u_{k}^{\ast}(t)
&=
\langle \phi^{-}_{k}(t)|\rho_{0,k}(t)|\phi^{+}_{k}(t)\rangle,
\\
v_{k}^{\ast}(t)u_{k}(t)
&=\langle \phi^{+}_{k}(t)|\rho_{0,k}(t)|\phi^{-}_{k}(t)\rangle. 
\el
\ee
%
%
In the presence of noise, the averaged density matrix $\rho_{k}(t)$ for a given mode $k$ and white-noise realization $R(t)$
are obtained by numerically solving the exact master equation 
for  \cite{Budini2000,Filho2017,Kiely2021,Jafari2024,Baghran2024,Sadeghizade2025,Asadian2025},
%
\be
{d\over dt}\rho_{k}(t)=-i[{\cal H}_{0,k}(t),\rho_{k}(t)]-\frac{\xi^2}{2}[{\cal H}_1,[{\cal H}_1,\rho_{k}(t)]],
\label{eq:master}
\ee
%
where ${\cal H}_{0,k}(t)$ is the noise-free Hamiltonian while $R(t) {\cal H}_1= R(t) \sigma^{z}$ expresses 
the ``noisy" part for the full Hamiltonian ${\cal H}_{\xi,k}(t)={\cal H}_{0,k}(t)+R(t){\cal H}_1$. This master equation has the form of a von Neumann equation with an added term $-({\xi^2}/{2})[\sigma^z , [\sigma^z,\rho_{k}(t)]]$ representing the effect from the noise.


Next, we evaluate the quantum Fisher information and Wigner–Yanase skew information for our system. Owing to translational invariance under periodic boundary conditions, it suffices to consider the two‐spin reduced density matrix $\varrho_{\ell,\ell+r}(t)$ for an arbitrary pair of sites $\ell$ and $\ell+r$. In particular,  any such two‐qubit density matrix can be written in the form  
%
\bea
\label{eq4}
\varrho_{\ell,\ell+r}(t)=\left(\begin{array}{cccc}
 \rho_{11} &0 &0 & \rho_{14} \\
 0 & \rho_{22} & \rho_{23} &0 \\
 0 & \rho_{23}^{\ast} & \rho_{22} &0 \\
  \rho_{14}^{\ast} &0 &0 &\rho_{44} \\
\end{array}\right)\, ,
\eea
%
where the matrix elements can be written in terms of one- and two-point correlation functions (see Appendix. \ref{appA} for details).

\section{Quantum Fisher Information}
Estimation theory is an important topic in different areas of physics \cite{Chin2012,Mishra2021,Montenegro2021,Alipour2015,Rezakhani2019}.
In general phase estimation perspective, the evolution of a mixed quantum state, given by the density matrix $\varrho$,
under a unitary transformation, can be described as $\varrho_\theta=e^{-i A\theta}\varrho e^{i A\theta}$, 
where $\theta$ is the phase shift and $A$ is an operator.
The estimation accuracy for $\theta$ is bounded by the quantum Cram\'{e}r-Rao inequality \cite{Helstrom,Holevo}:
%
\be
\Delta\hat{\theta}\ge\frac{1}{\sqrt{\nu F(\varrho_\theta)}},
\ee
%
%
where $\hat{\theta}$ expresses the unbiased estimator for $\theta$, $\nu$ is the number of times 
the measurement is repeated, and $F(\varrho_\theta)$ is the so-called the quantum Fisher information. It  is given by~\cite{Helstrom,Holevo}
%
\begin{equation}
\label{eq17}
	F(\varrho_\theta, {\cal  O} )=2\sum_{m,n}\frac{(p_m-p_n)^2}{(p_m+p_n)}|\langle m| {\cal O} |n\rangle|^2,
\end{equation}
%
where $p_m$ and $|m\rangle$ denote  the eigenvalues and eigenvectors of the
density matrix $\varrho_\theta$, respectively, which is used as a probe state to estimate $\theta$.
Here, ${\cal O}$ is the observable with respect to whom $F$ needs to be optimized.
Moreover,  for an arbitrary bipartite state one can show~\cite{Yin2019,Li2013}
%
\begin{equation}
\label{eq18}
F=\sum_\mu F(\varrho_\theta, A_\mu\otimes I+I\otimes B_\mu),
\end{equation}
%
where $\{A_\mu\}$ and $\{B_\mu\}$ are arbitrary and natural complete sets of local orthonormal observables of the two subsystems that $\varrho_\theta$ is composed of.
For a general two-spin system, the local orthonormal observables $\{A_\mu\}$ and $\{B_\mu\}$ can be chosen as \cite{Lei2016}
%
\be
\{A_\mu\}=\{B_\mu\}=\frac{1}{\sqrt{2}}\{I, \sigma^x, \sigma^y, \sigma^z\}.
\ee
%
Consequently, for a given $\varrho_\theta$ is given, $F$ can be calculated from Eq. (\ref{eq17}).
The reduced two-spin density matrix, Eq.~(\ref{eq4}), facilitates the analytical evaluation of QFI
of the two-spin density matrix. 
Thus,  the eigenvalues and their corresponding normalized eigenvectors of the density
matrix   can be  obtained as 
%
\be
\bl
\label{11}
&
p_{1}=\frac{1}{2}(\rho_{11}+\rho_{44}+\sqrt{(\rho_{11}-\rho_{44})^{2}+4|\rho_{14}|^{2}}),\\
&
p_{2}=\frac{1}{2}(\rho_{11}+\rho_{44}-\sqrt{(\rho_{11}-\rho_{44})^{2}+4|\rho_{14}|^{2}}),\\
&
p_{3}=\frac{1}{2}(\rho_{22}+\sqrt{4|\rho_{23}|^{2}});
\quad
p_{4}=\frac{1}{2}(\rho_{22}-\sqrt{4|\rho_{23}|^{2}}),
\el
\ee
%
and
%
\be
\bl
\label{eq12}
&
|\phi _{1}\rangle=N_{1}\left(
                       \begin{array}{c}
                         \rho_{14} \\
                         0 \\
                         0 \\
                         p_{1}-\rho_{11} \\
                       \end{array}
                     \right),
                     \quad
|\phi _{2}\rangle=N_{2}\left(
                       \begin{array}{c}
                         \rho_{14} \\
                         0 \\
                         0 \\
                         p_{2}-\rho_{11} \\
                       \end{array}
                     \right),
                     \\
&
|\phi _{3}\rangle=N_{3}\left(
                       \begin{array}{c}
                         0 \\
                         \rho_{23} \\
                          p_{3}-\rho_{22} \\
                         0 \\
                       \end{array}
                     \right),
                     \quad
|\phi _{4}\rangle=N_{4}\left(
                       \begin{array}{c}
                         0 \\
                         \rho_{23} \\
                          p_{4}-\rho_{22} \\
                         0 \\
                       \end{array}
                     \right),
\el
\ee
%
in which $N_{i}$ ($i = 1,2,3,4$) are the normalization factors.
Then, the analytical evaluation of the QFI can be evaluated as
%
\begin{eqnarray}
\no
F&=&\frac{(p_4-p_2)^2}{p_2+p_2}\Big(\frac{\omega_{-}\chi_{+}^2+1}{q\chi_{+}-1}\Big) - \frac{(p_4-p_1)^2}{p_4+p_1} \Big(\frac{\omega_{-}\chi_{-}^2+1}{q\chi_{-}-1}\Big)\\
\no
&-& \frac{(p_3-p_2)^2}{p_3+p_2}\Big(\frac{\omega_{+}\chi_{+}^2+1}{q\chi_{+}-1}\Big) + \frac{(p_3-p_1)^2}{p_3+p_1}\Big(\frac{\omega_{+}\chi_{-}^2+1}{q\chi_{-}-1}\Big)\\
\label{eq19}
&+& \frac{(p_2-p_1)^2}{p_2+p_1}\Big(\frac{1}{(\chi_{+}^2+1)(\chi_{-}^2+1)}\Big),
\end{eqnarray}
%
where $\omega_{\pm}=(\sqrt{\rho_{23}}\pm\sqrt{\rho_{23}})^2/4|\rho_{23}|$, $\chi_{\pm}=\rho_{44}-\rho_{11}\pm\sqrt{(\rho_{44}-\rho_{11})^2+4|\rho_{14}|^2}$, 
and $q=(\rho_{11}-\rho_{44})/2|\rho_{14}|$.

\section{Wigner-Yanase skew information}{\label{WYSI}}
To assess how noisy ramp quenches influence spin–spin correlations in our XY model, we also consider WYSI as a measure of collective coherence. 
Formally, the WYSI is defined as~\cite{Wigner1963,Girolami2014,Karpat2014,CAKMAK2012,Jafari2020a,Lei2016}
%
\bea
I(\varrho,V)=-\frac{1}{2}
 {\rm Tr}
[\sqrt{\varrho},V]^{2},
\eea
%
where the density matrix $\varrho$ depict a mixed quantum state and $V$ is an
observable.
The quantity $I(\varrho,V)$ can also be interpreted as a measure of the quantum uncertainty of
$V$ in the state $\varrho$ instead of the conventional variance.
A set of the local spin elements ($\sigma^{\alpha}_\ell$) is an arbitrary and natural choice of observable which constitutes a local orthonormal basis.
Hence, we employ the local quantum coherence (LQC) as
%
\bea
{\rm LQC}^\alpha_{\ell,\ell+r}= I(\varrho_{\ell,\ell+r},\sigma^{\alpha}_{\ell}\otimes \openone_{\ell+r}).
\eea
%
%
By straightforward calculations, the root of the two-qubit reduced state $\sqrt{\varrho_{\ell,\ell+r}}$
can be obtained by
%
\bea
\label{eq14}
\sqrt{\varrho_{\ell,\ell+r}}=\left(
                       \begin{array}{cccc}
                         \alpha_\varrho & 0 & 0 & \lambda_\varrho \\
                         0 & \beta_\varrho & \nu_\varrho & 0 \\
                         0 & \nu_\varrho^{\ast} & \gamma_\varrho & 0 \\
                         \lambda_\varrho^{\ast} & 0 & 0 & \delta_\varrho \\
                       \end{array}
                     \right),
\eea
%
with the following elements
%
\be
\bl
\label{eq15}
\alpha_\varrho
=&
|\rho_{14}|^{2}\Big(\frac{\sqrt{p_{1}}}{N_{1}^{2}}+\frac{\sqrt{p_{2}}}{N_{2}^{2}}\Big);
\quad
\beta_\varrho
=
|\rho_{23}|^{2}\Big(\frac{\sqrt{p_{3}}}{N_{3}^{2}}+\frac{\sqrt{p_{4}}}{N_{4}^{2}}\Big),
\\
\gamma_\varrho
=&
\frac{\sqrt{p_{3}}(p_{3}-\rho_{22})^{2}}{N_{3}^{2}}+\frac{\sqrt{p_{4}}(p_{4}-\rho_{22})^{2}}{N_{4}^{2}},
\\
\delta_\varrho
=&
\frac{\sqrt{p_{1}}(p_{1}-\rho_{11})^{2}}{N_{1}^{2}}+\frac{\sqrt{p_{2}}(p_{2}-\rho_{11})^{2}}{N_{2}^{2}},
\\
\lambda_\varrho
=&
\rho_{14}\Big(\frac{\sqrt{p_{1}}(p_{1}-\rho_{11})}{N_{1}^{2}}+\frac{\sqrt{p_{2}}(p_{2}-\rho_{11})}{N_{2}^{2}}\Big),
\\
\nu_\varrho
=&
\rho_{23}\Big(\frac{\sqrt{p_{3}}(p_{3}-\rho_{11})}{N_{3}^{2}}+\frac{\sqrt{p_{4}}(p_{4}-\rho_{11})}{N_{4}^{2}}\Big).
\el
\ee
%
For a bipartite system in the form of Eq.~(\ref{eq4}), the two-spin LQC components can be written as~\cite{Lei2016}
%
\be
\bl
\label{eq16}
{\rm LQC}^x_{\ell,\ell+r}
&=
1-2(\alpha_\varrho \beta_\varrho +\gamma_\varrho \delta_\varrho )-4{\mathrm Re}
\Big[
\lambda_\varrho \nu_\varrho
\Big],
\\
{\rm LQC}^y_{\ell,\ell+r}
&=
1-2(\alpha_\varrho \beta_\varrho +\gamma_\varrho \delta_\varrho)+4
{\mathrm Re}
\Big[
\lambda_\varrho \nu_\varrho
\Big],
\\
{\rm LQC}^z_{\ell,\ell+r}
&=
1-
\Big[
\alpha_\varrho^{2}+\beta_\varrho^{2}+\gamma_\varrho^{2}+\delta_\varrho^{2}-2
\Big(
|\lambda_\varrho|^{2}+|\nu_\varrho|^{2}
\Big)
\Big],
\el
\ee
%
which quantify the coherence with respect to the first subsystem locally.

\section{Numerical Results}

In this section, we illustrate the foregoing theoretical framework through comprehensive numerical simulations of the transverse‐field Ising model (setting $\gamma=1$), highlighting the dynamical behavior of both Quantum Fisher Information and Wigner-Yanas-Skew Information under noiseless and noisy ramp protocols.

%
\begin{figure}[t!]
\centering
\includegraphics[width=0.495\linewidth]{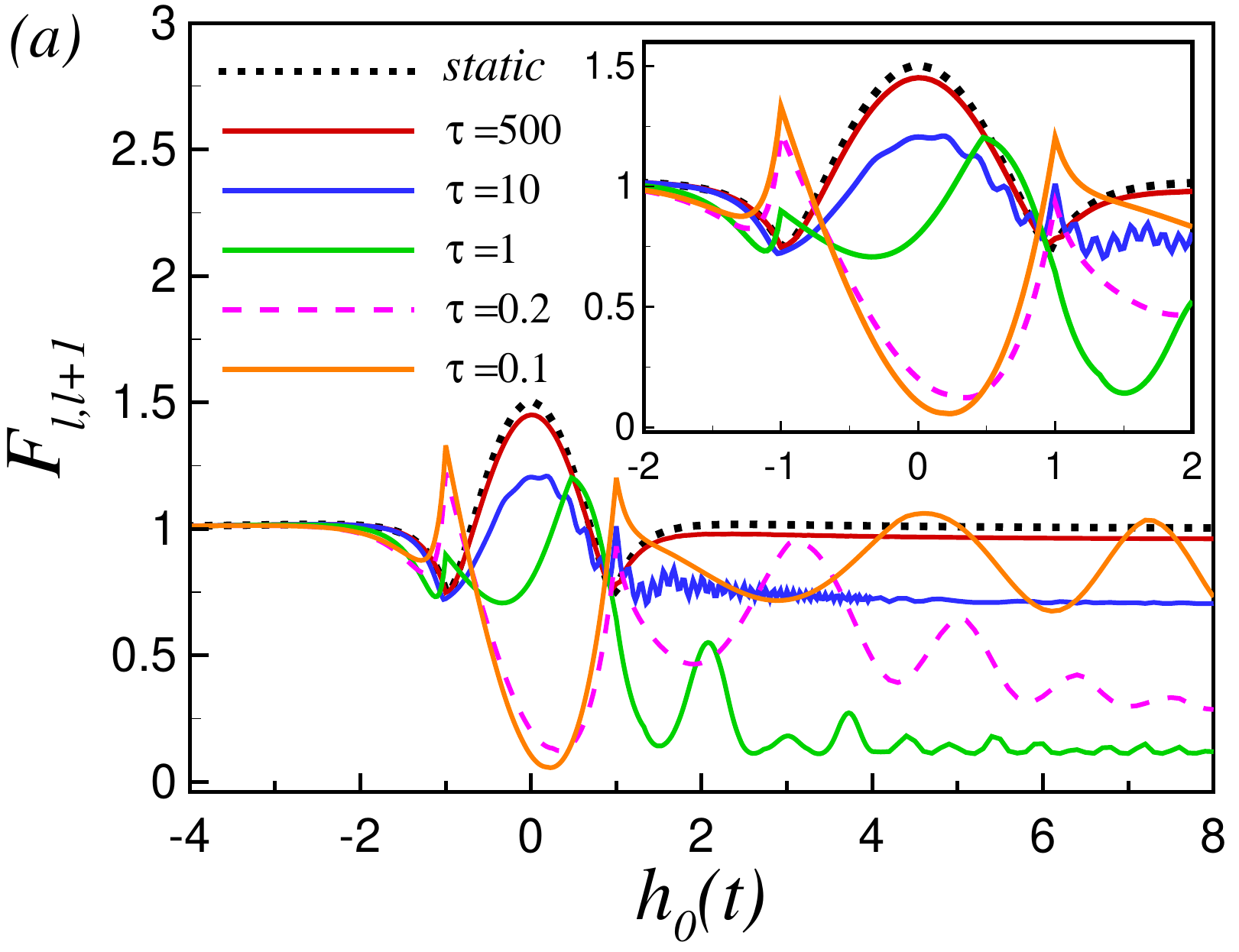}%
\hfill
\includegraphics[width=0.495\linewidth]{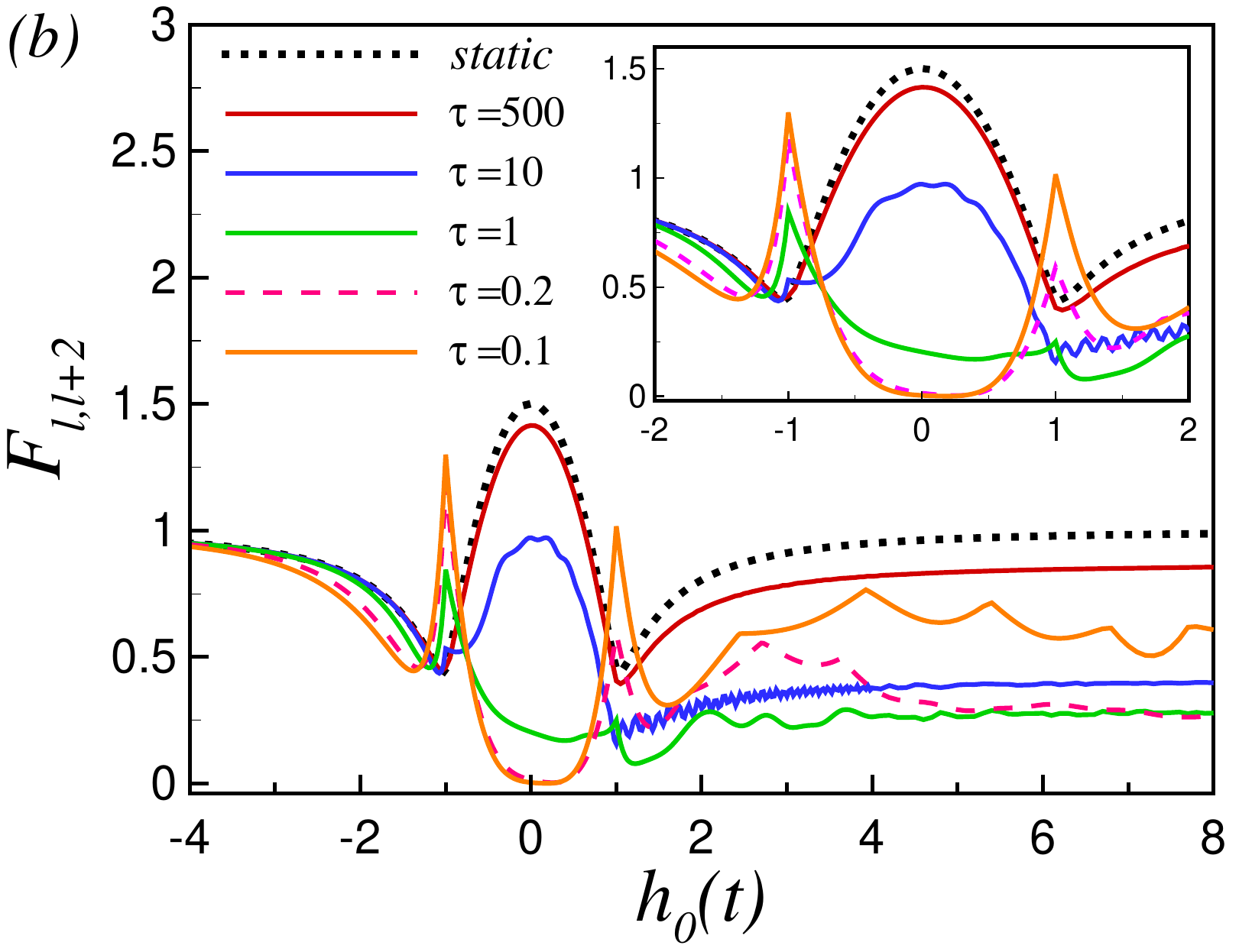}
\vspace{1em}
\includegraphics[width=0.495\linewidth]{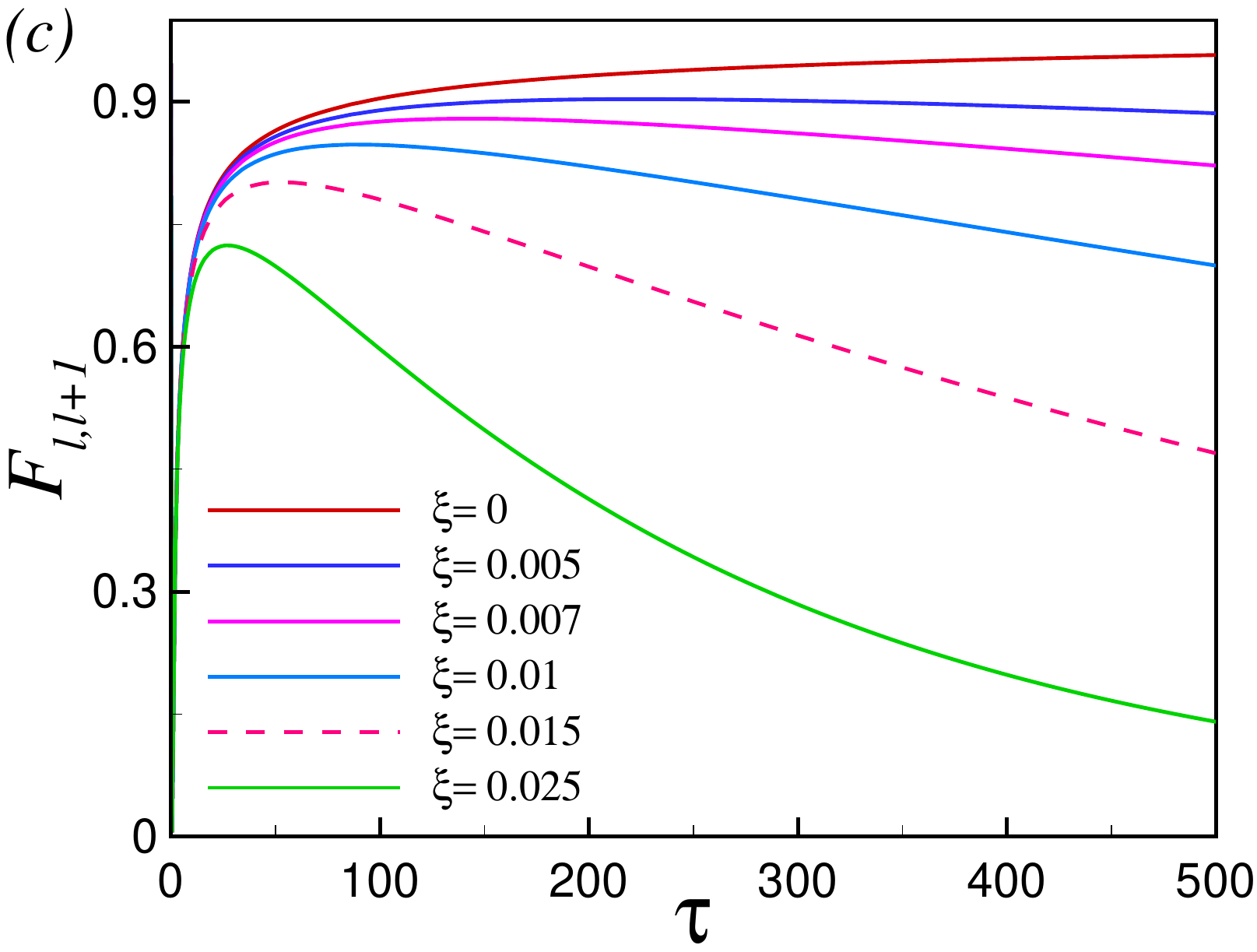}%
\hfill
\includegraphics[width=0.495\linewidth]{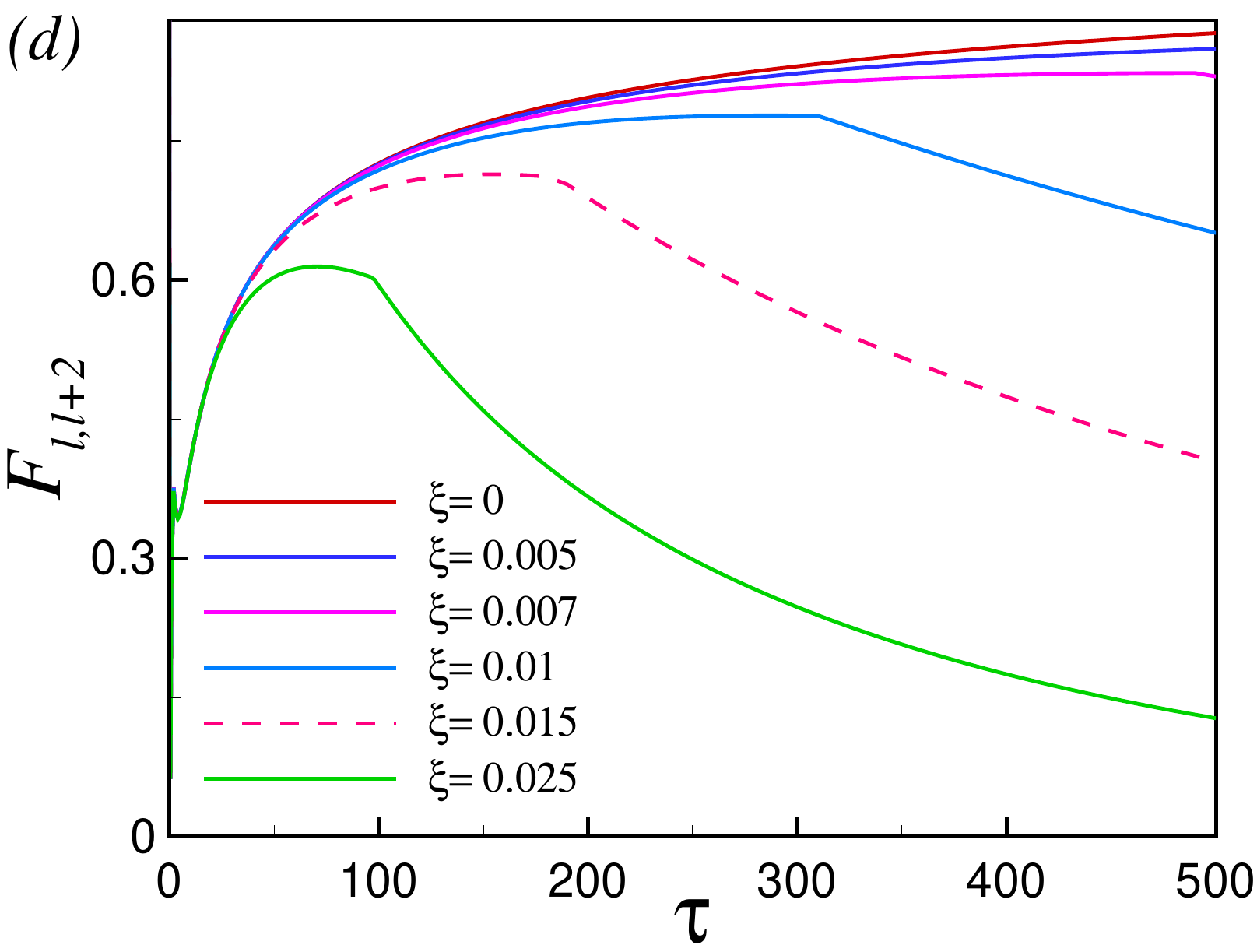}
\caption{
Quantum Fisher information for nearest‐neighbour $F_{\ell,\ell+1}$ (a), and next‐nearest‐neighbour $F_{\ell,\ell+2}$ (b) as a function of the instantaneous field $h(t)$ during a linear ramp from $h_i=-30$ to $h_f$, in the absence of noise. Insets zoom into the critical region $-2<h(t)<2$.  
The lower panel show the  Quantum Fisher information $F_{\ell,\ell+1}$ (c) and $F_{\ell,\ell+2}$ (d) versus ramp duration $\tau$ for a full quench from $h_i=-30$ to $h_f=30$.
}
\label{fig1}
\end{figure}

%
\begin{figure*}[t!]
\centerline{
\includegraphics[width=0.28\linewidth]{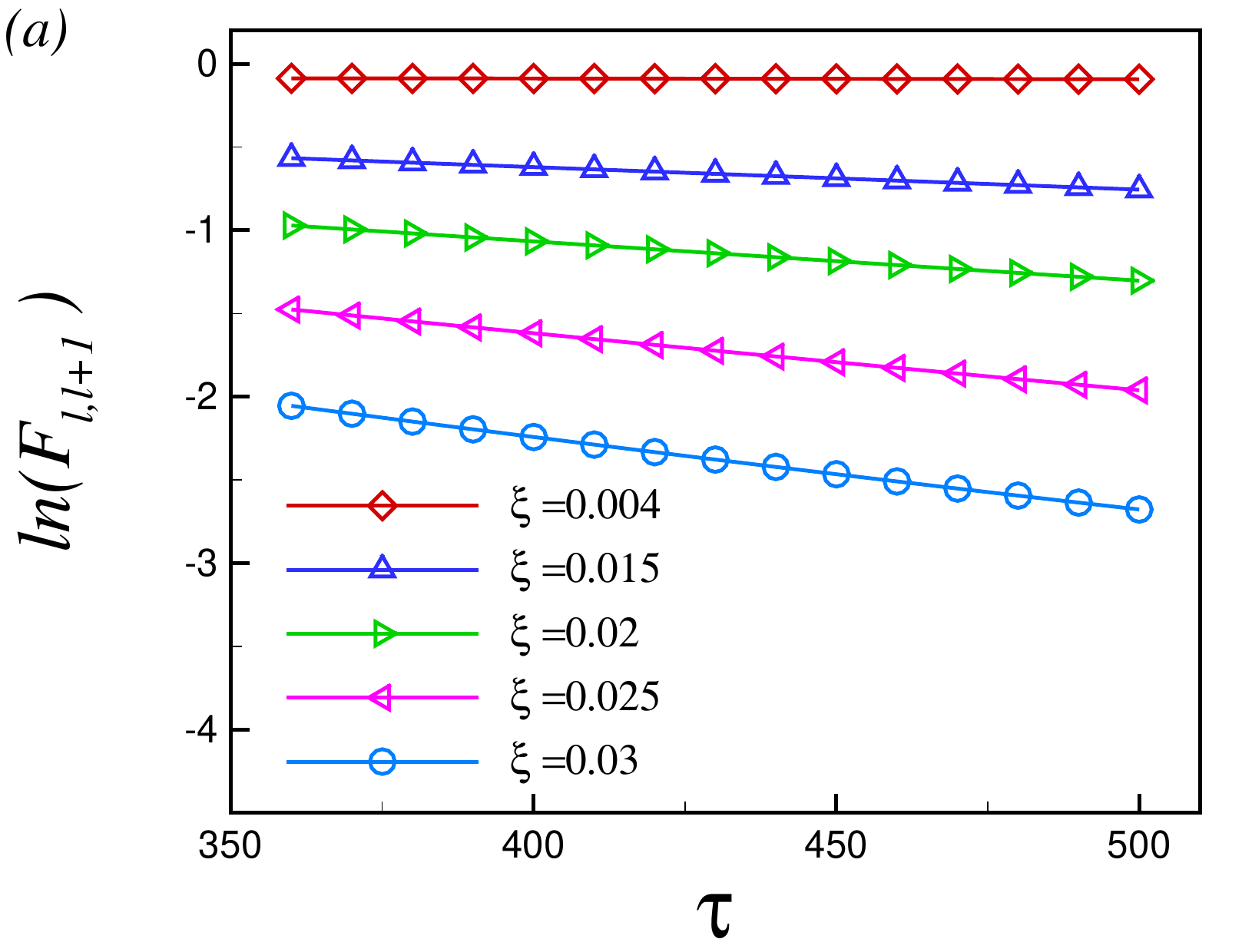}
\includegraphics[width=0.28\linewidth]{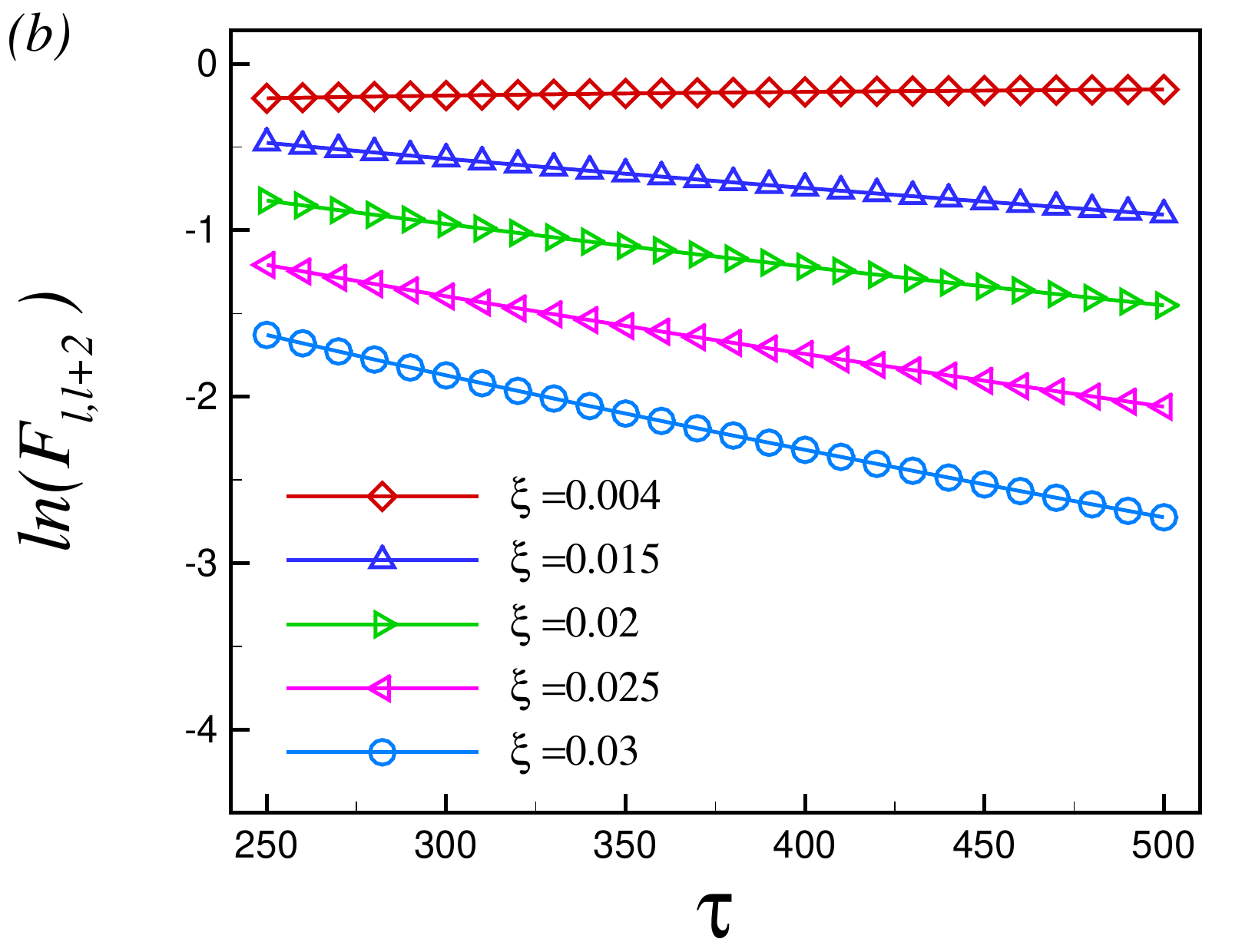}
\includegraphics[width=0.28\linewidth]{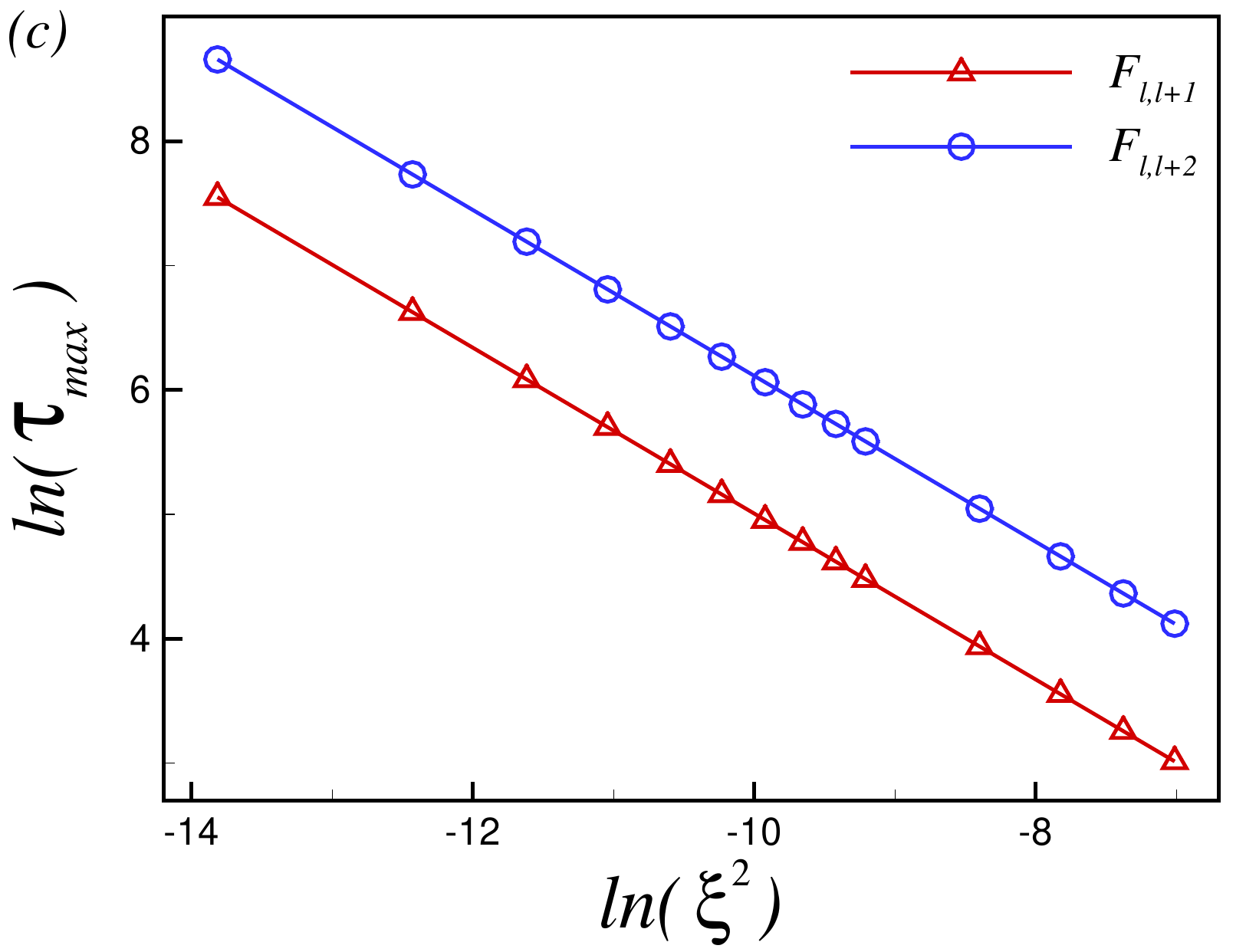}}
\caption{
 The variation of (a) $\ln(F_{l,l+1})$ and 
(b) $\ln(F_{l,l+1})$ as a function of $\tau$ for the noisy quench 
from $h_i=-30$ to $h_f=30$ for different values of noise intensity $\xi$, 
reveals the linear scaling of logarithm of quantum Fisher information versus $\tau$.
(c) The optimal time of $F_{\ell,\ell+1}$ and $F_{\ell,\ell+2}$ are shown to have power-law scaling with square of the strength of noise $\xi^2$ with exponent $\delta=0.66\pm0.02$.}
\label{fig2}
\end{figure*}
%

%
\begin{figure*}[t]
\begin{minipage}{\linewidth}
\centerline{
\includegraphics[width=0.28\linewidth]{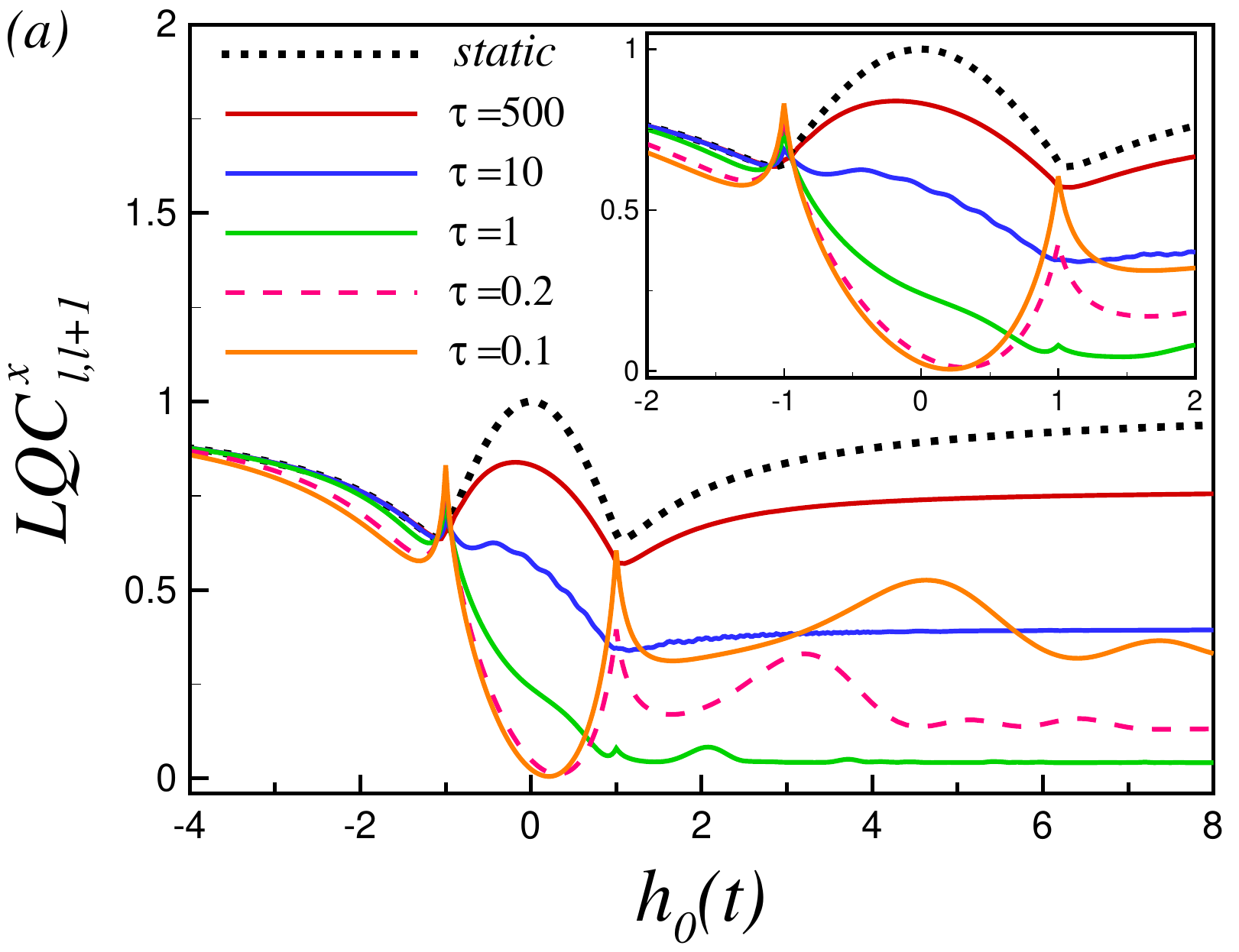}
\includegraphics[width=0.28\linewidth]{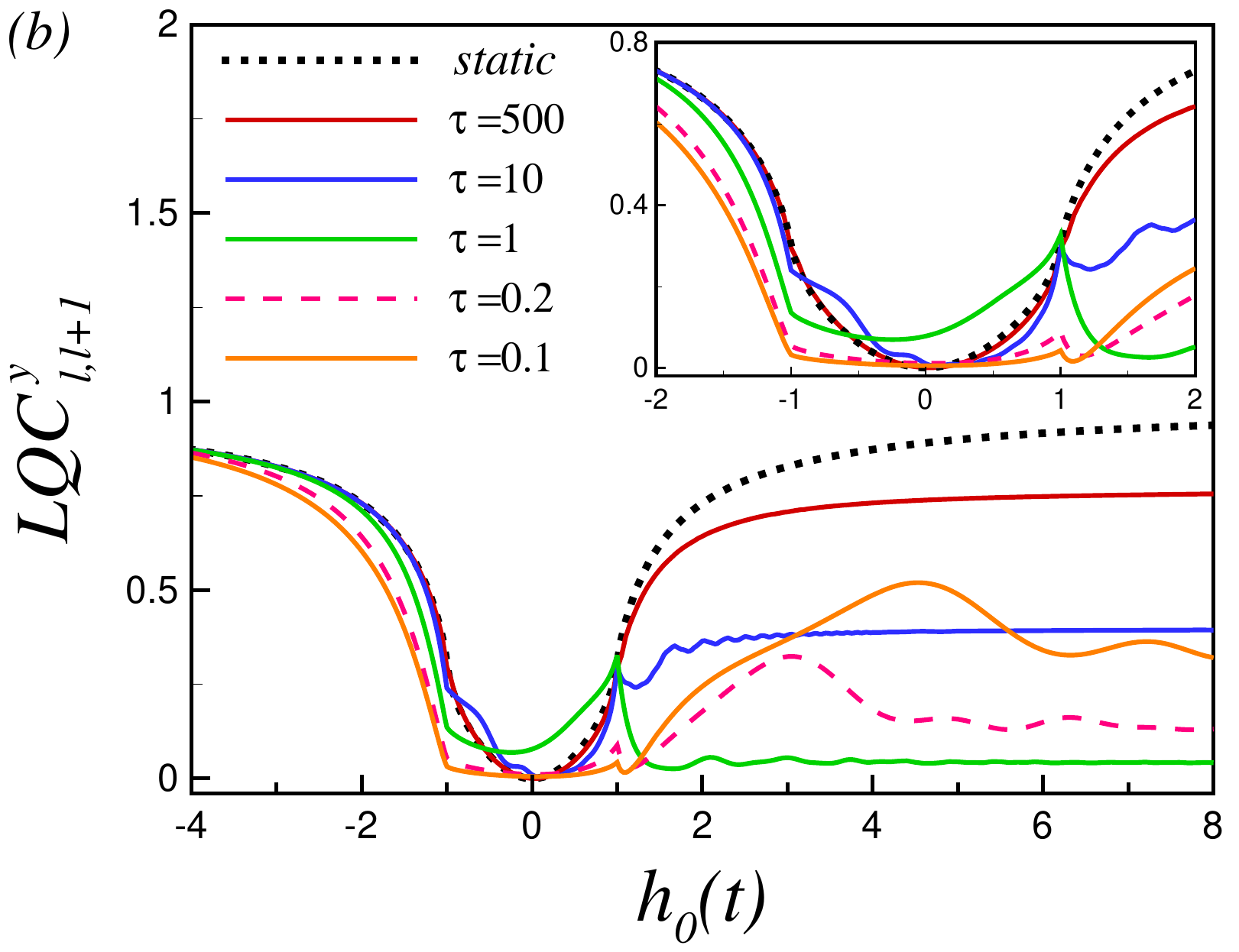}
\includegraphics[width=0.28\linewidth]{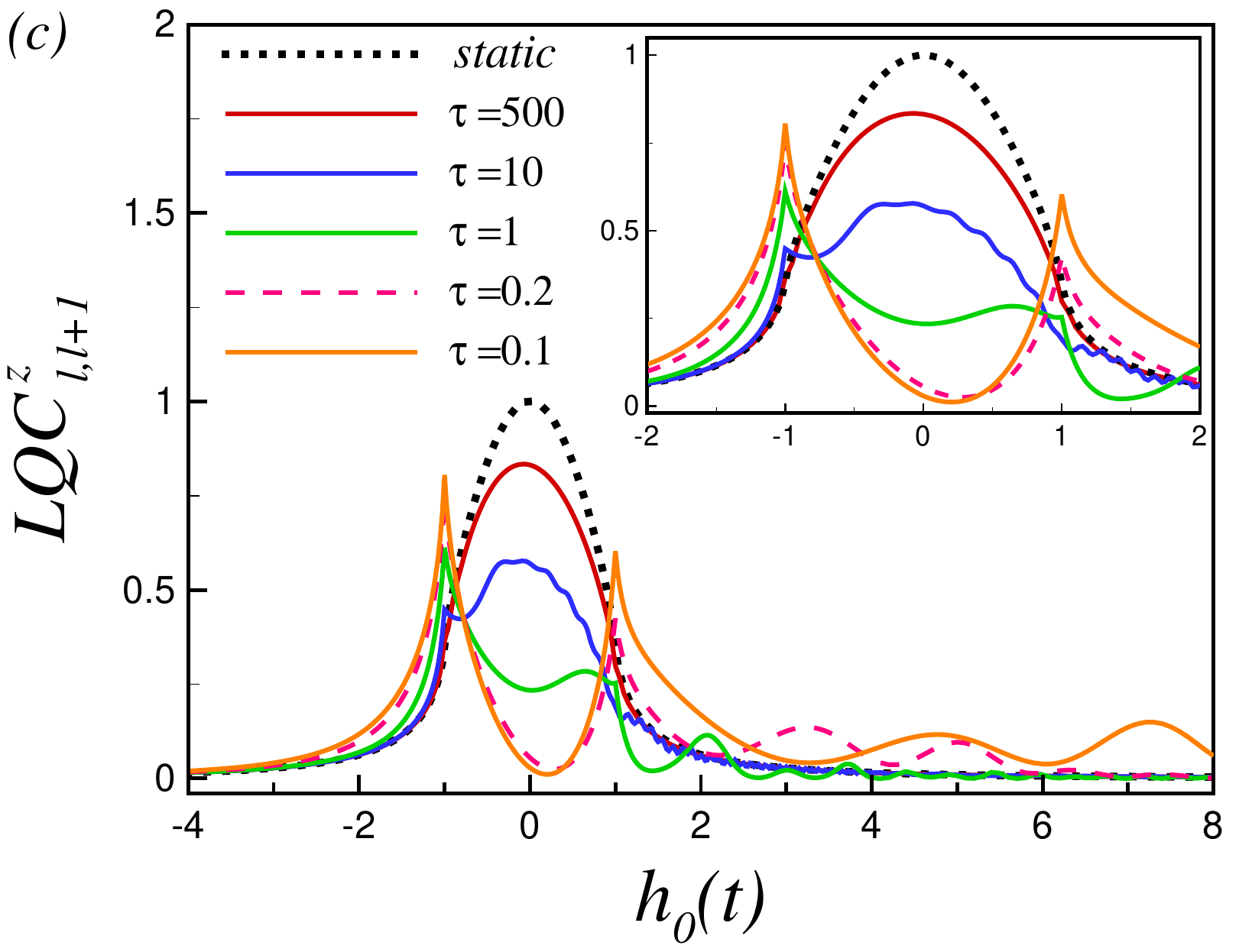}}
\centering
\end{minipage}
\begin{minipage}{\linewidth}
\centerline{
\includegraphics[width=0.28\linewidth]{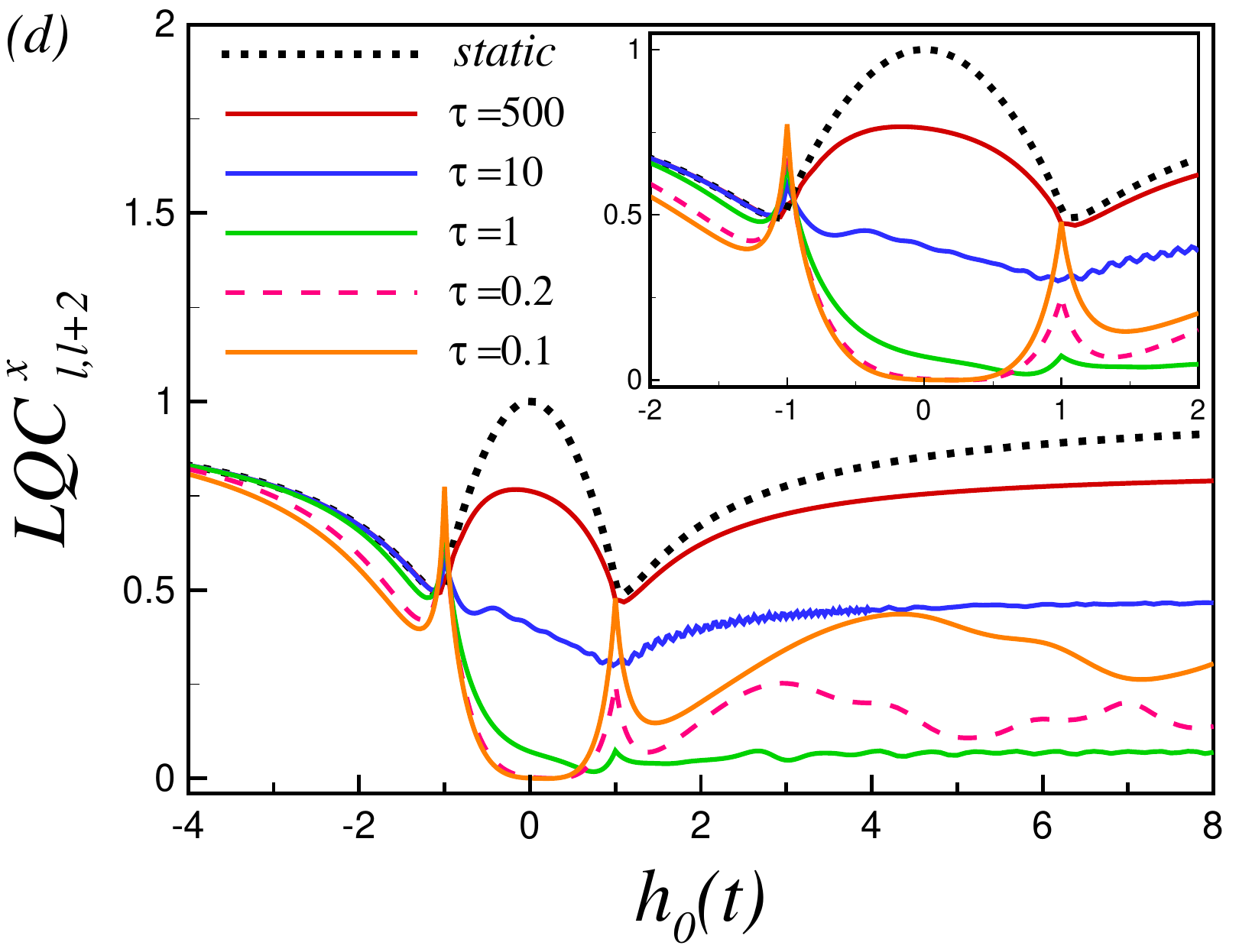}
\includegraphics[width=0.28\linewidth]{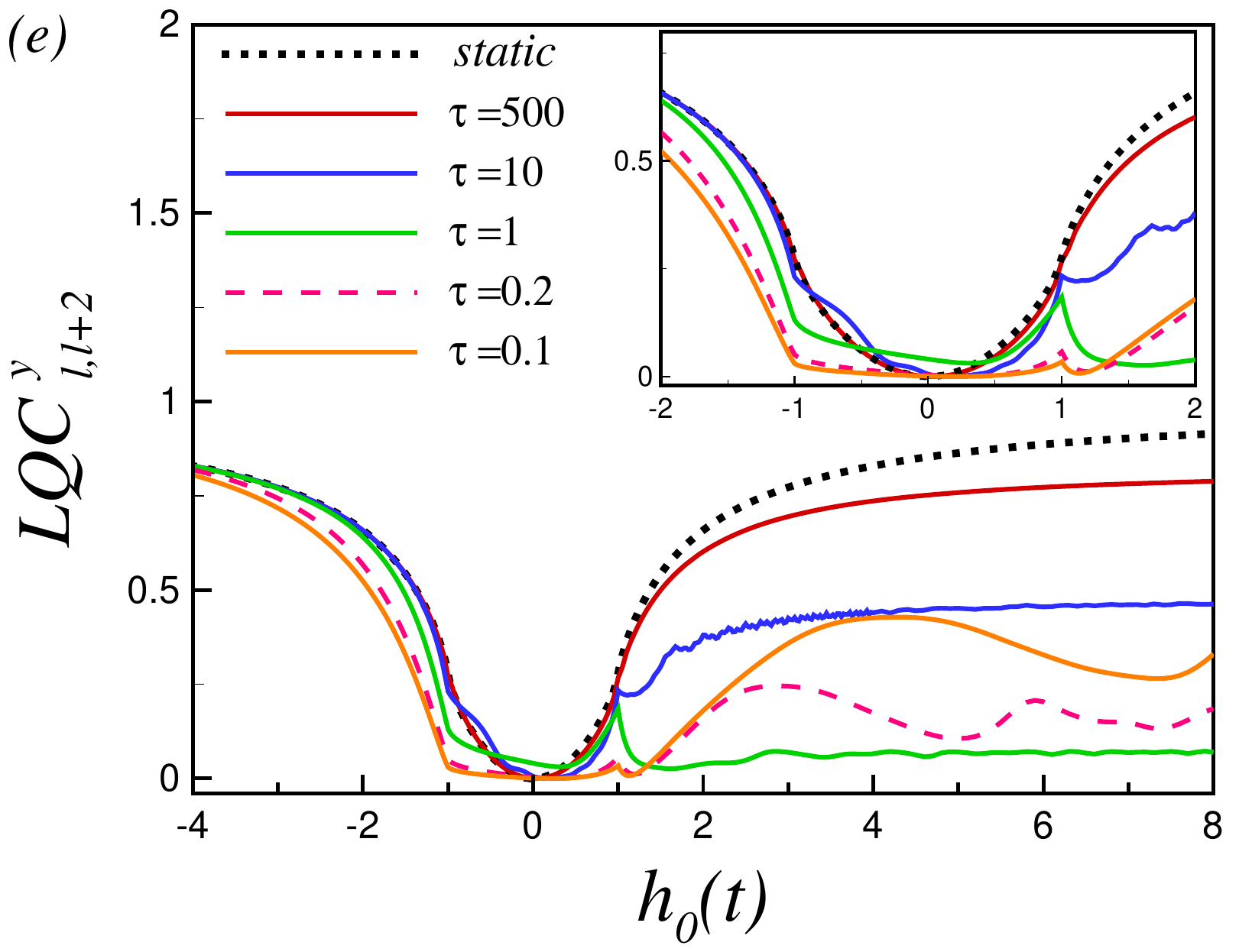}
\includegraphics[width=0.28\linewidth]{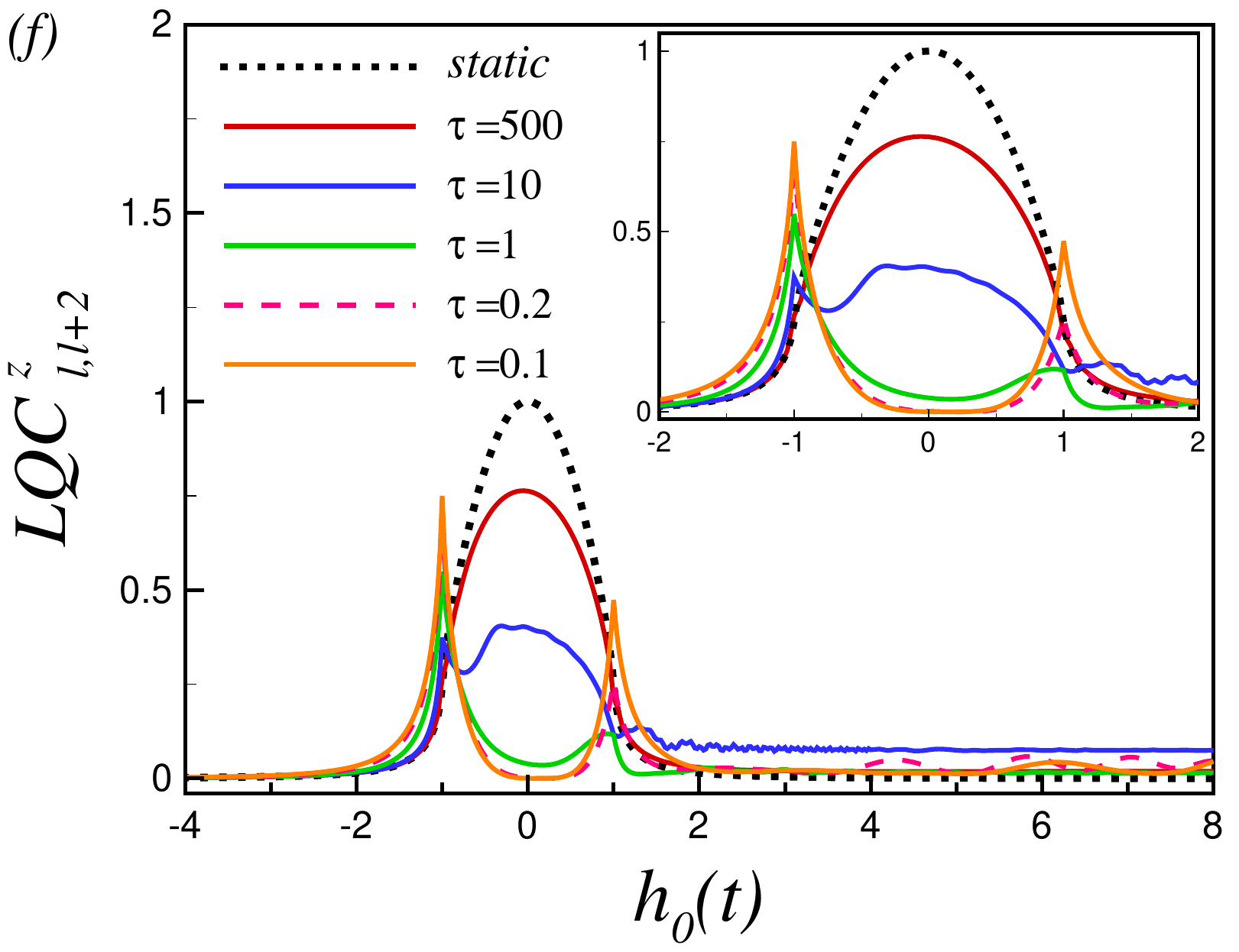}}
\centering
\end{minipage}
\caption{
(a)-(c) Wigner–Yanase skew information between nearest‐neighbour spins ${\rm LQC}^{\alpha}_{\ell,\ell+1}$ for $\alpha=\{x,y,z\}$ as a function of the instantaneous field $h(t)$ during a noiseless ramp from $h_i=-30$ to $h(t)$.
(d)-(f) Wigner–Yanase skew information for next‐nearest‐neighbour spins ${\rm LQC}^{\alpha}_{\ell,\ell+2}$ under the same ramp protocol and noiseless conditions.  
 Insets zoom into the critical region $-2<h(t)<2$.  
}
\label{fig3}%
\end{figure*}%
%

\begin{figure}[t!]
  \centering
  \includegraphics[width=0.495\linewidth]{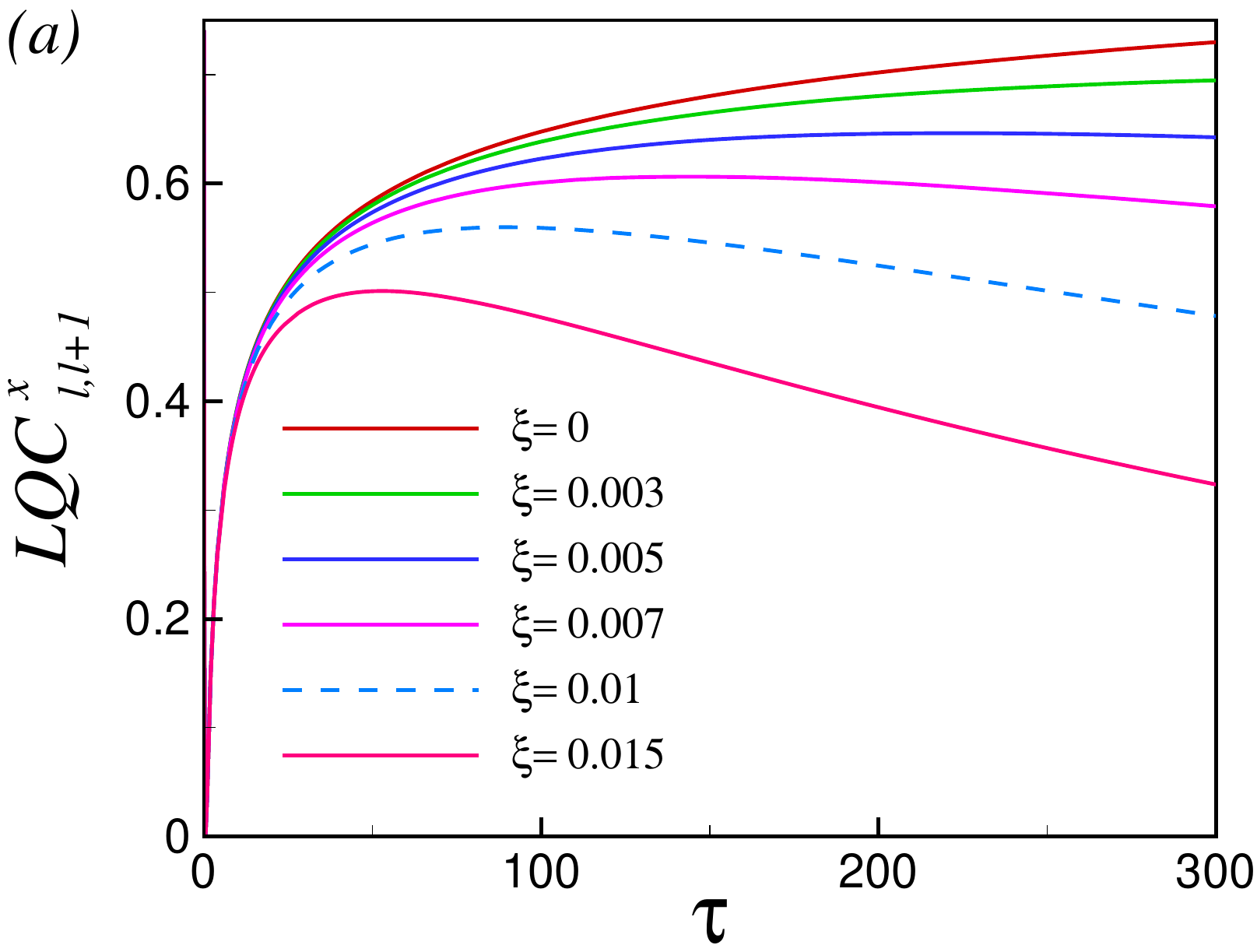}\hfill
  \includegraphics[width=0.495\linewidth]{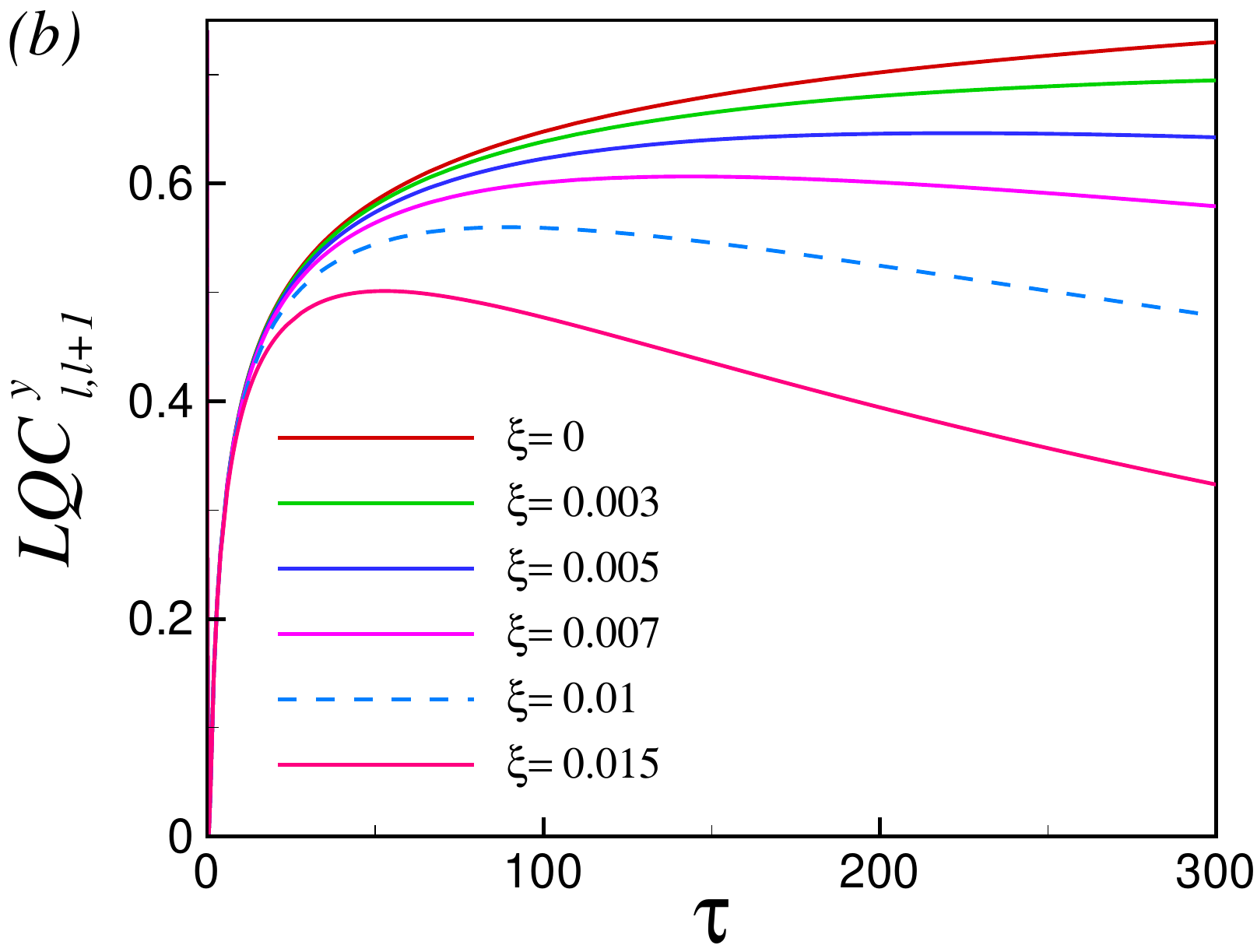}

  \vspace{0.5em}

  \includegraphics[width=0.495\linewidth]{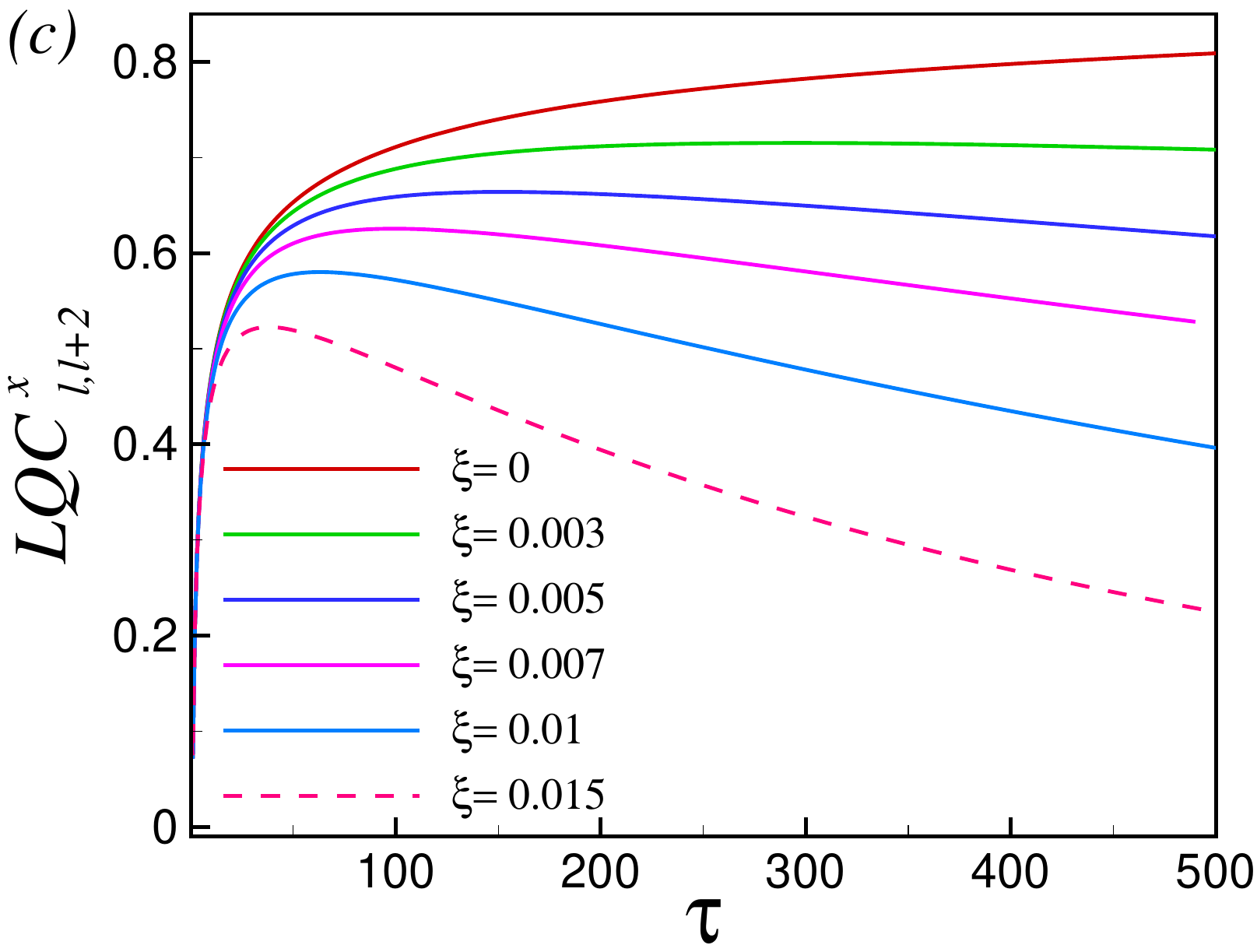}\hfill
  \includegraphics[width=0.495\linewidth]{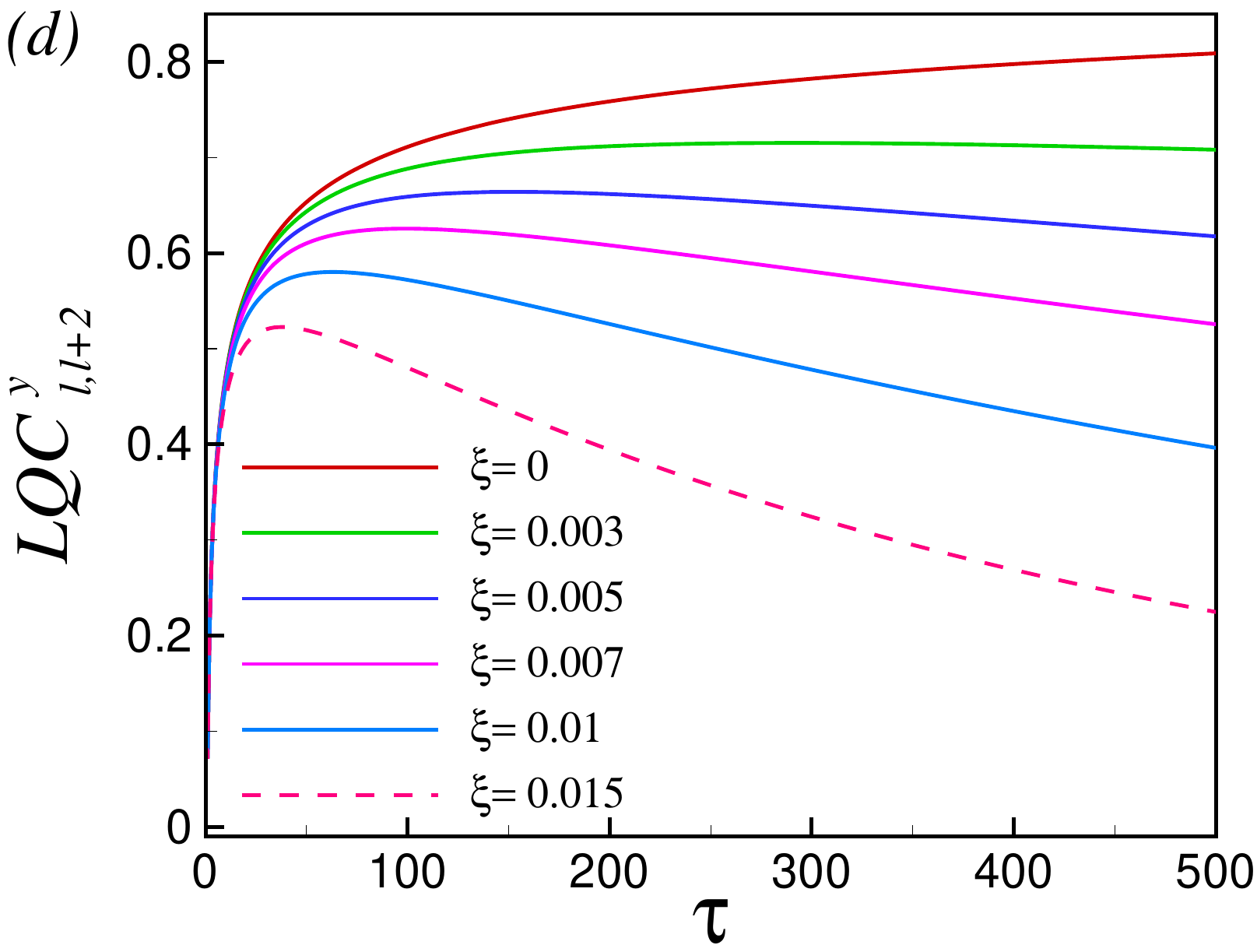}

  \caption{
  Wigner–Yanase skew information versus ramp duration $\tau$ for a quench from $h_i=-30$ to $h_f=30$: (a) ${\rm LQC}^{x}_{l,l+1}$, (b) ${\rm LQC}^{y}_{l,l+1}$, (c) ${\rm LQC}^{x}_{l,l+2}$, and (d) ${\rm LQC}^{y}_{l,l+2}$.
  }
  \label{fig4}
\end{figure}

\begin{figure}[t!]
  \centering
  \includegraphics[width=0.495\linewidth]{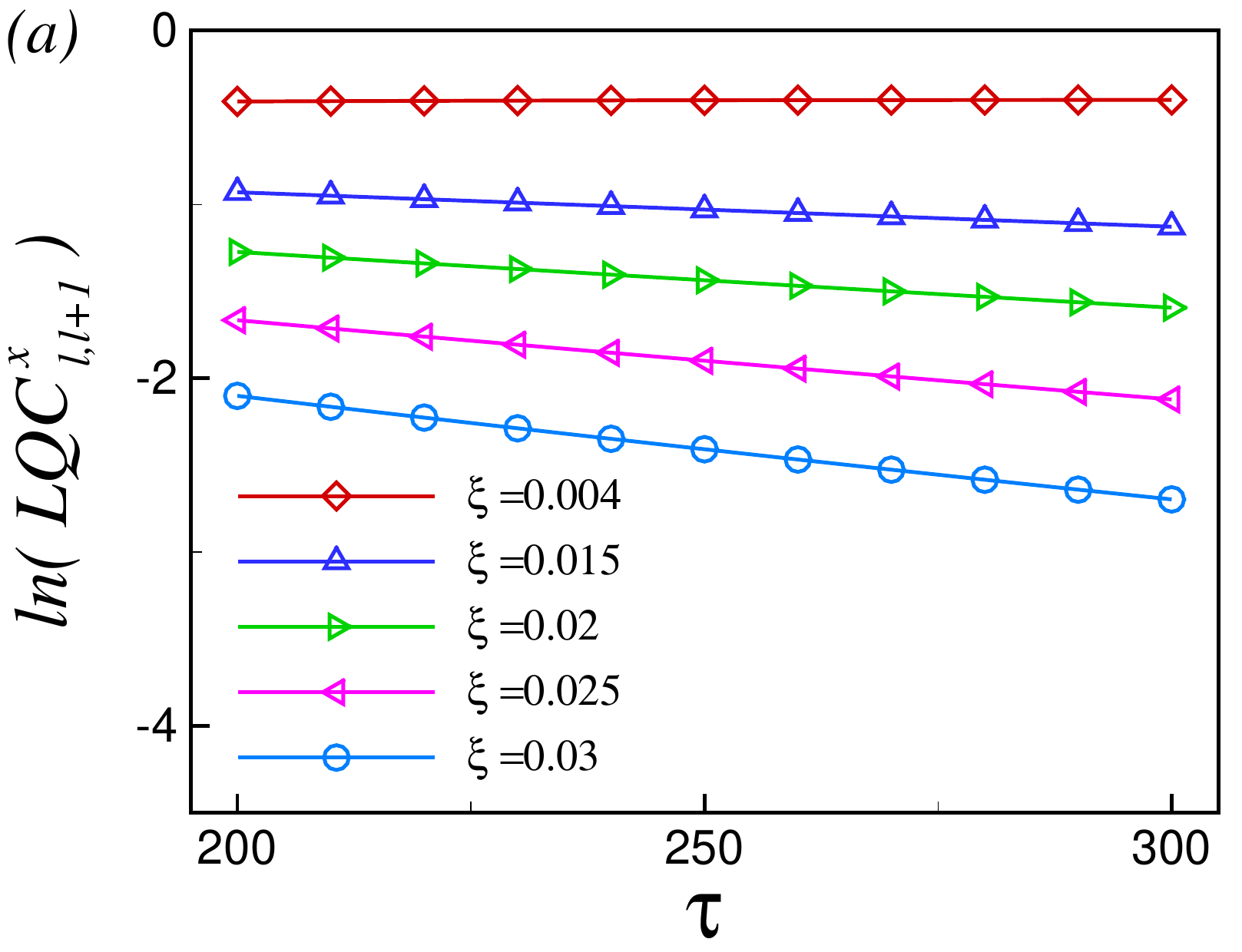}\hfill
  \includegraphics[width=0.495\linewidth]{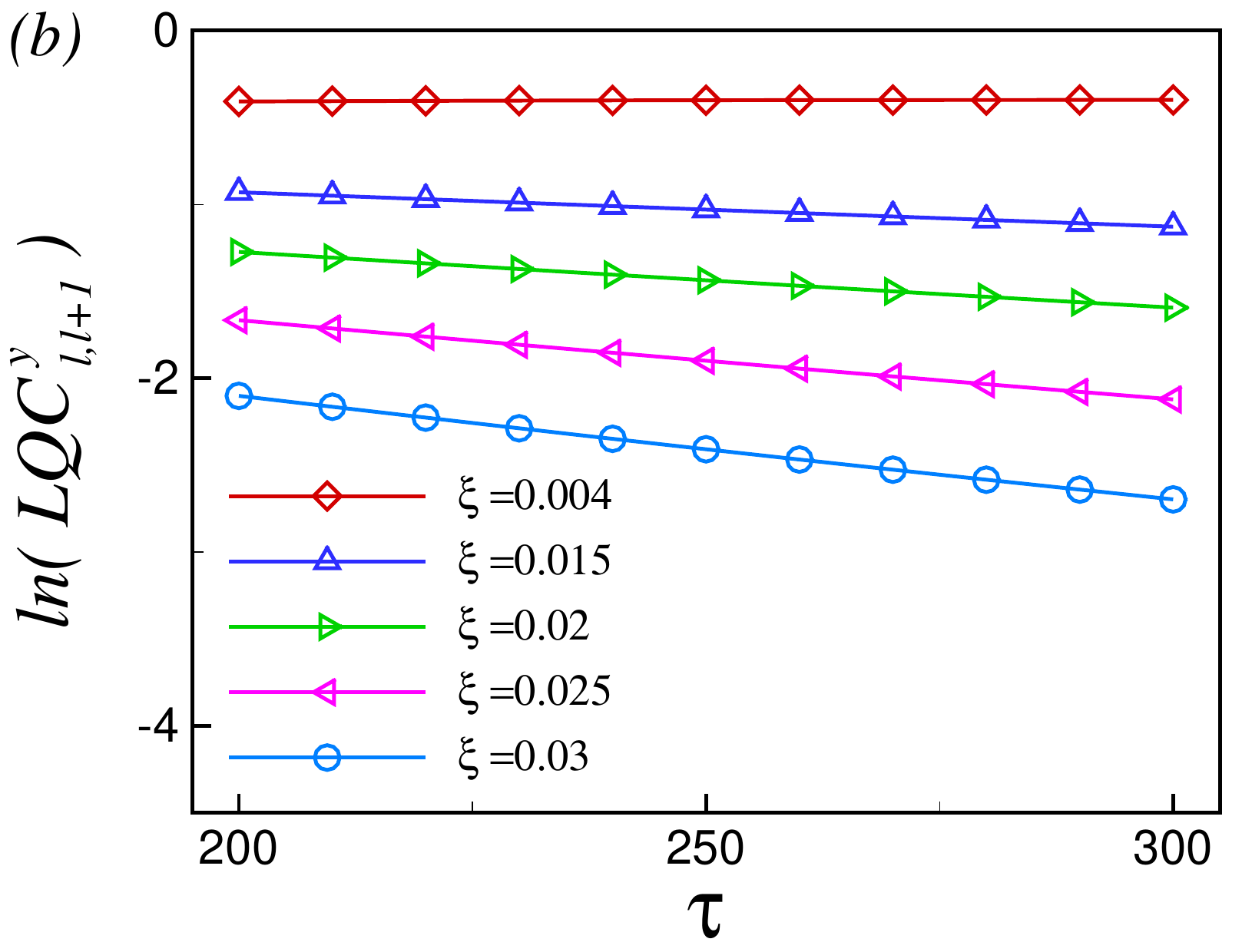}

  \vspace{0.5em}
  \includegraphics[width=0.495\linewidth]{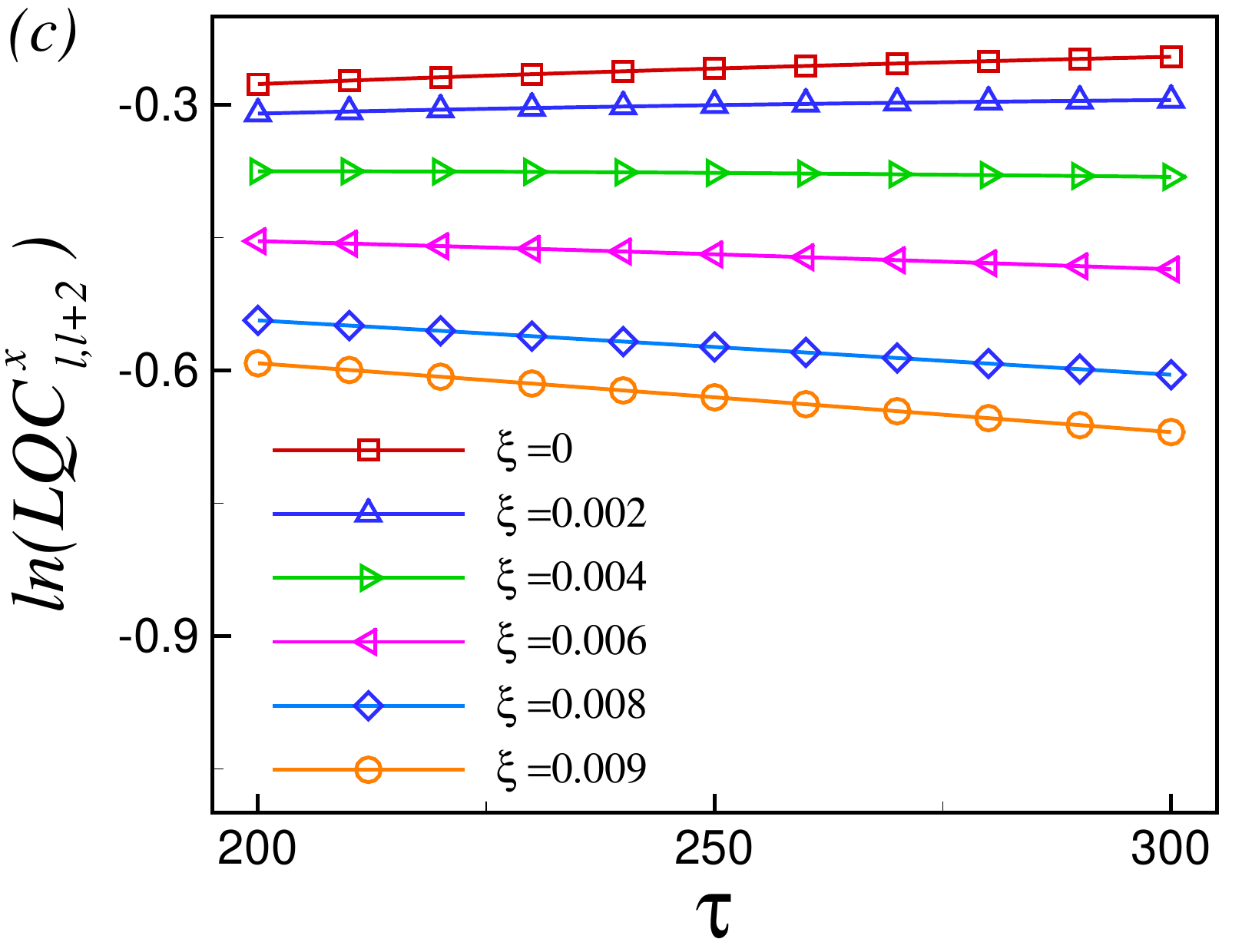}\hfill
  \includegraphics[width=0.495\linewidth]{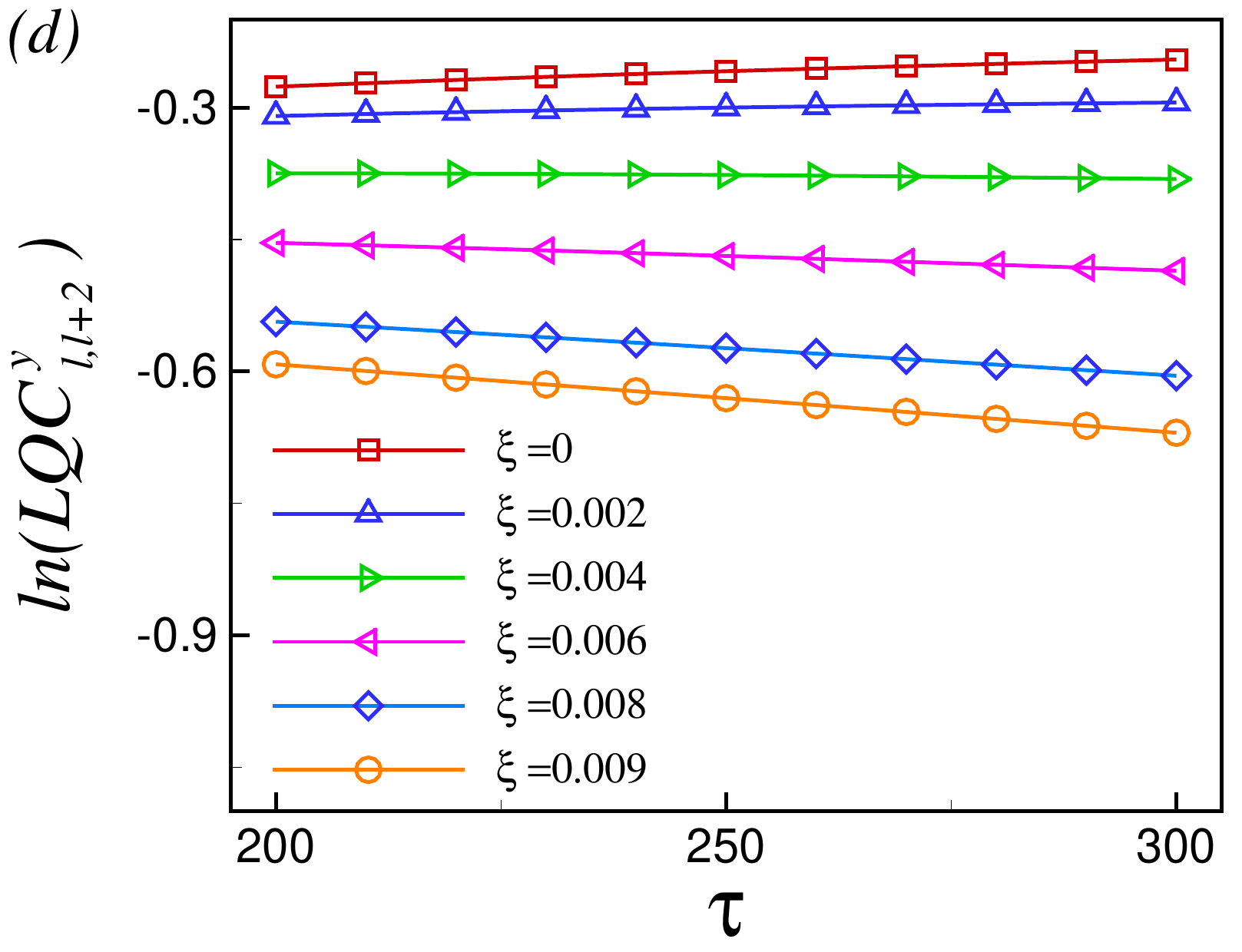}

  \vspace{0.5em}
  \includegraphics[width=0.55\columnwidth]{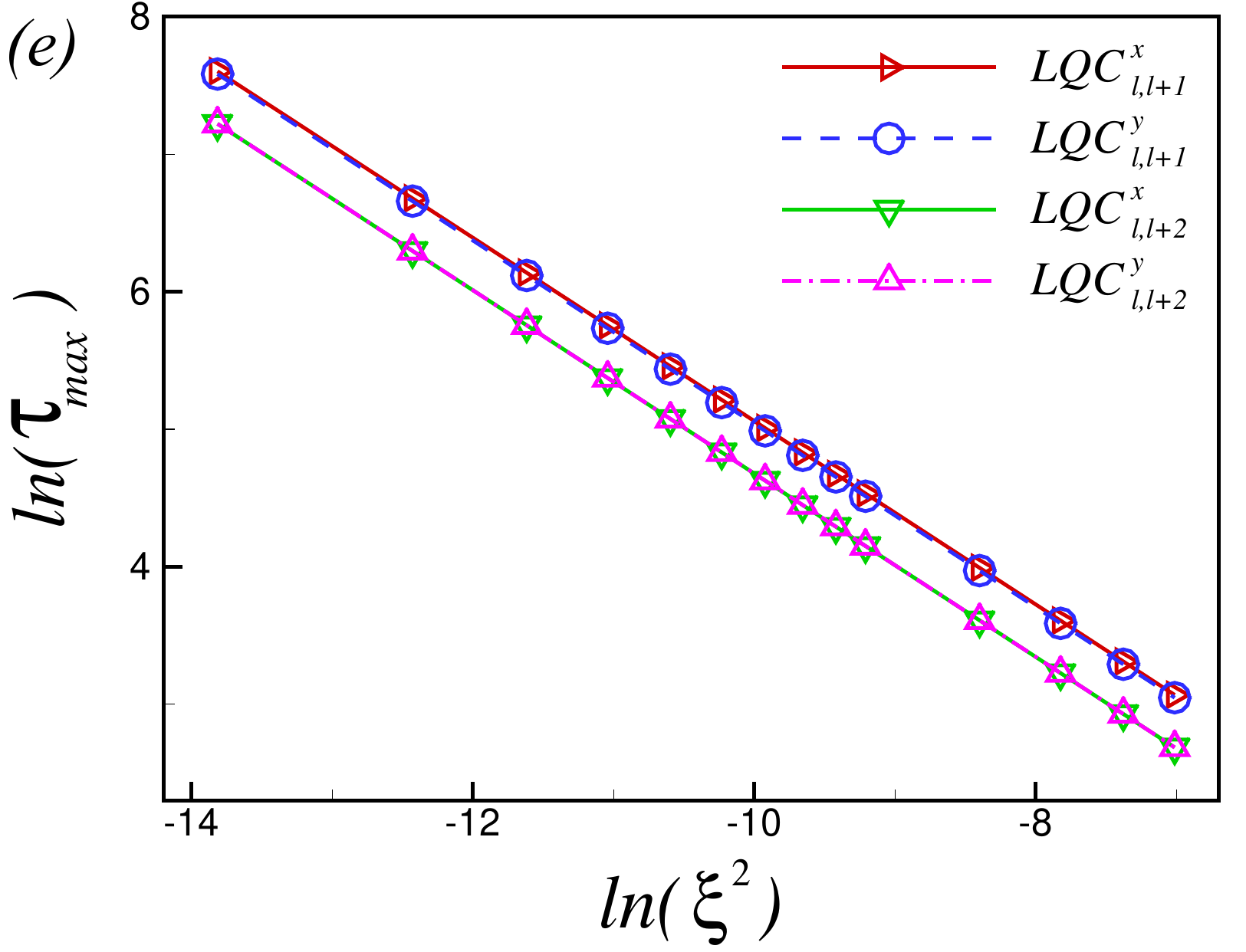}

  \caption{Logarithm of Wigner–Yanase skew information versus ramp duration $\tau$ for a noisy quench from $h_i=-30$ to $h_f=30$ at various noise intensities $\xi$: (a) $\ln({\rm LQC}^{x}_{l,l+1})$, (b) $\ln({\rm LQC}^{y}_{l,l+1})$, (c) $\ln({\rm LQC}^{x}_{l,l+2})$, (d) $\ln({\rm LQC}^{y}_{l,l+2})$, demonstrating a linear scaling.
   (e) 
  The $\tau$ that  maximises   ${\rm LQC}^{\alpha}_{\ell,\ell+1}$ and ${\rm LQC}^{\alpha}_{\ell,\ell+2}$, $\tau_m$, shows power-law scaling with respect to the square of the noise strength $\xi^2$, with the exponent $\delta = 0.66 \pm 0.02 \approx 2/3$.}
  \label{fig5}
\end{figure}

\subsection{Quantum Fisher Information}
In our numerical investigation, we first explore the dynamics of the local quantum Fisher information  under both noiseless and noisy linear quenches. Figures~\ref{fig1}(a) and (b) show $F_{\ell,\ell+1}$ and $F_{\ell,\ell+2}$, respectively, as functions of the instantaneous field $h_0(t)$ for a chain of length $N=200$, with initial field $h_i=-30$ and several ramp durations $\tau$. 
In contrast to the dynamics of entanglement in anisotropic XY chains, local quantum Fisher information is non-zero between nearest-neighbor spins.
During the adiabatic evolution (very slow quench $\tau=500$), QFI closely follow the static values, i.e., those obtained from the ground 
state of the system at each value of $h_0(t)$, and assumes its minimum values at the critical points $h_c=\pm1$, see Fig.~\ref{fig1}(a) and (b).
This behavior is expected from the adiabatic theorem \cite{Kato1950,messiah1999} and as a consequence of the 
finite size of the system. Indeed the energy gap closes as an inverse function of the system size, remaining nonzero for any finite size $L$, so that it is always possible to reach the adiabatic limit provided $\tau$ is long enough.
More precisely, it has been shown that, the probability of having an adiabatic evolution at size
$L$ is given by $p(\tau)=1-\exp(-2\pi^3\tau/L^2)$  \cite{Cincio2007}, so that the maximum quench rate at which the evolution is adiabatic 
decays as $1/L^2$.
By decreasing $\tau$ the quench time to intermediate values ($10<\tau<20$), QFI is nonzero at each value of $h_0(t)$, with small oscillations.

As the time scale of the ramping decreases ($0.5\le\tau\lesssim10$) the QFI oscillations become larger and faster as the system crosses the first critical point $h_c{=}-1$.
By further reduction of the quench time ($\tau\lesssim 0.5$) both $F_{l,l+1}$ and $F_{l,l+2}$ reveal maximum at the critical point $h_c{=}-1$ and then decays gradually to reach the minimum around $h(t)=0$.
The QFI then increases with $h(t)$ till reaching its maximum value at the second critical point $h_c=1$, and gradually decays afterwards with oscillations around the mean value and finally equilibrates to mean value at large $h_0(t)$.
The maximum value of QFI at the critical point is the result of an optimal mixing of all eigenstates in the system and oscillatory behavior has been ascribed to the fact that the system ends up, after passing the second critical point, in a superposition of excited states of the instantaneous Hamiltonian \cite{Calabrese2005}. While in the static case the critical points are signaled by the minimum of the QFI.
For very large values of $\tau$, i.e., very slow quenches: the period of the oscillation and the time over which QFI oscillates around the mean value to equilibrate are enhanced.
To better understand the effect of noise on the dynamics of QFI, we have plotted $F_{l,l+1}$ and $F_{l,l+2}$ versus $\tau$ in Fig.~\ref{fig1}(c)-(d) for a quench from 
$h_i=-30$ to $h_f=30$ in the absence and presence of the noise. As seen, in the noiseless case $\xi=0$, both $F_{l,l+1}$ and $F_{l,l+2}$ increases by enhancing $\tau$ which 
is in contrast to the behavior of entanglement and defect density \cite{Cherng2006,Anirban2016,singh2021,Jafari2025a}. While in the presence of the noise $F_{l,l+1}$ and $F_{l,l+2}$ increases monotonically with $\tau$ and reaches a maximum value at an intermediate $\tau_m(\xi)$ and starts decreasing for $\tau>\tau_m(\xi)$.
As the noise intensity $\xi$ increase $\tau_m(\xi)$ is also shifted to the lower value and magnitude of QFI decreases. The numerical analysis shows that, QFI scales exponentially 
with $\tau$ with negative noise-dependent exponent, i.e., $F_{l,l+r} \propto \tau^{-\upnu_\xi}$ with $r=1$ and $r=2$ as illustrated in Fig.~\ref{fig2}(a)-(b).
Moreover, a more detailed analysis shows that the quench time scale at which QFI is maximum, i.e., $\tau_m(\xi)$ scales with square of noise intensity as $\tau_m(\xi)\propto(\xi^2)^{-\delta}$ with the exponent $\delta=0.66\pm0.02\approx 2/3$ which has been illustrated in Fig.~\ref{fig2}(c). 
It should be mentioned that, this exponent  $\delta=2/3$ is the same as that which governs the scaling
of the optimal time for minimal defect production in a standard quantum annealing scheme \cite{Anirban2016,Gao2017}.


\subsection{Wigner-Yanas-Skew Information}
We now turn to the WYSI features under both noiseless and noisy quenches. As in our previous analysis of the quantum Fisher information, we compute the WYSI between nearest-neighbour spins, ${\rm LQC}^{\alpha=\{x,y,z\}}_{\ell,\ell+1}$, and next-nearest-neighbour spins, ${\rm LQC}^{\alpha}_{\ell,\ell+2}$. Figs.~\ref{fig3}(a)-(c) display the WYSI as a function of the instantaneous field $h_0(t)$ for various ramp times $\tau$ in the nearest-neighbour case, while Figs.~\ref{fig3}(d)-(f) show the corresponding results for next-nearest-neighbours. All data were obtained for a ramp from $h_i=-30$ to $h(t)$  
 in the absence of noise.
Notably, Fig.~\ref{fig3}(a) shows that the nearest‐neighbour component ${\rm LQC}^x_{\ell,\ell+1}$ exhibits almost identical behaviour to the quantum Fisher information (Fig.~\ref{fig2}(a)). More precisely, as $\tau$ increases (i.e., slower quenches), the curves converge towards the static (adiabatic) limit. At $h_0(t)=0$, a pronounced maximum hump appears for slow quenches, but as the quench rate increases (smaller $\tau$), this central peak broadens, its amplitude decreases, and eventually turns into a minimum. Simultaneously, the minima observed at $h_0(t)=\pm1$ in the slow‐quench regime invert into maxima under fast quenches.
On the other hand, the WYSI component ${\rm LQC}^y_{\ell,\ell+1}$ shows markedly different behaviour (Fig.~\ref{fig3}(b)). Even in the static limit, it displays a clear minimum at $h_0(t)=0$ such as the  fast quenches. While a pronounced maximum persists at $h_0(t)=1$, the feature at $h_0(t)=-1$ becomes only a kink rather than a true peak.
Turning to Fig.~\ref{fig3}(c), ${\rm LQC}^z_{\ell,\ell+1}$ mirrors the $x$–component’s approach to adiabaticity, with a pronounced maximum at $h_0(t)=0$ for slow quenches, the static limit, underscoring the dominant longitudinal correlations at zero field. As the quench rate increases, this central peak broadens and its amplitude diminishes, yet it persists as a maximum rather than inverting into a minimum. Furthermore, at high fields ($|h_0(t)|\gg1$), ${\rm LQC}^z_{\ell,\ell+1}$ decays to zero, in stark contrast to the finite plateaus reached by the $x$ and $y$ components.
Moreover, the next‐nearest‐neighbour component ${\rm LQC}^\alpha_{\ell,\ell+2}$ (Fig.~\ref{fig3}(d)-(f)) closely mirrors the behaviour of its nearest‐neighbour counterpart. In Fig.~\ref{fig3}(d)–(f) one again observes pronounced features at the critical fields: a central extremum at $h_0(t)=0$ that broadens and diminishes with increasing quench rate, inflections or secondary extrema near $h_0(t)=\pm1$, and, for the $z$-component, a decay toward zero at large $|h_0(t)|$. Compared to the nearest‐neighbour curves, all peaks and plateaus are slightly reduced in amplitude, reflecting the weaker coherence between spins separated by two lattice sites.

To compare with our previous QFI results, we plot in Figs.~\ref{fig4}(a)-(d), the Wigner–Yanase skew information versus ramp duration $\tau$ for a quench from $h_i=-30$ to $h_f=30$ for both nearest‐ and next‐nearest‐neighbour $x$ and $y$ components. We omit the $z$‐component here, as it is negligible at high fields and would add little to this analysis.
In the noiseless scenario, the skew information increases monotonically with $\tau$, rising rapidly at small $\tau$ due to strong nonadiabatic excitations and then saturating to its static (adiabatic) value for large $\tau$. This behaviour holds for all four cases: nearest‐neighbour $x$ and $y$, and next‐nearest‐neighbour $x$ and $y$.
In the presence of noise, however, the skew information no longer simply saturates. Instead, analogous to the QFI, each curve develops a maximum at an intermediate $\tau_m(\xi)$ and then decreases for $\tau>\tau_m(\xi)$.
Thus, all four $x$ and $y$ components, whether nearest‐ or next‐nearest neighbours, display qualitatively the same approach to QFI, with the next‐nearest‐neighbour curves slightly attenuated in amplitude faster  by increasing $\tau$, for higher noise intensity.

Finally, in Figs.~\ref{fig5}(a)-(d), we plot $\ln ({\rm LQC}^{\alpha=x,\; y}_{\ell,\ell+r})$ versus ramp duration $\tau$ for a noisy quench from $h_i=-30$ to $h_f=30$ at various noise intensities $\xi$, where panels (a) and (b) correspond to the nearest‐neighbour  components and (c) and (d) to the next‐nearest‐neighbour counterparts. All four curves are strikingly linear over a wide range of $\tau$, confirming that the skew information decays exponentially with ramp time scale under noise. Moreover, extracting the $\tau_m$ at which each ${\rm LQC}^{\alpha}_{\ell,\ell+1}$ and ${\rm LQC}^{\alpha}_{\ell,\ell+2}$ reaches its peak and plotting $\ln\tau_m$ against $\xi^2$ (Fig.~\ref{fig5}(e)) reveals a power‐law scaling,  $\tau_m (\xi) \propto(\xi^{2})^{-\delta},$  
with exponent $\delta=0.66\pm0.02$, demonstrating that noise shifts the optimal ramp time the same as that of QFI.  
Is this a coincidence, or does it signify a relationship between defect formation and WYSI generation? If the latter is accurate, 
our finding of exponential scaling when noise is present still indicates that the two phenomena may not be as closely intertwined.

\section{Conclusions}

Motivated by the fundamental questions of how quantum correlations are generated under noisy quenches across quantum critical points and by the ubiquity of control‐field fluctuations in realistic implementations, we have examined the dynamics of quantum Fisher information (QFI) and Wigner–Yanase skew information (WYSI) in Ising chain subjected to noiseless and noisy linear driven transverse field.
In the absence of noise, and very large ramp time scale, the critical points $h_c=\pm 1$ are signaled by both QFI and ($x$) component of WYSI where become minimum.
While, as the ramp time scale decreases, both QFI and all component of WYSI represent cusps and the critical points. 
Moreover, both QFI and WYSI grow monotonically with the ramp time scale $\tau$, reflecting enhanced quantum coherence and metrological usefulness as the quench slows down, and saturate to their static (adiabatic) values for large $\tau$. 
This result is contrary to the behavior of entanglement between next-nearest-neighbor spins, which reveals suppression in accordance with $\tau^{-1/2}$ \cite{Cherng2006,Sengupta2009,Jafari2025a}.
%
When a Gaussian white noise of strength $\xi$ is added to the control field, the dynamics change qualitatively: both QFI and WYSI no longer simply saturate to the static values by increasing the ramp time scale, but instead develop a maximum at an intermediate ramp time scale $\tau_m$, beyond which they are suppressed by dephasing. By analyzing the logarithm of QFI and WYSI versus $\tau$, we demonstrated an exponential decay of both measures with ramp time scale under noise. 
While the density of defects in the presence of the noise does not scale exponentially with the ramp time scale \cite{Anirban2016,Gao2017}.
Moreover, the $\tau_m$ that maximizes each QFI and WYSI curve scales as
$
\tau_m(\xi) \propto (\xi^{2})^{\delta}, 
$
with a universal exponent $\delta \approx 2/3\,$ for both nearest‐ and next‐nearest‐neighbour spins.
This exponent corresponds to the exponent that dictates the scaling of the optimal time for achieving minimal defect production in a
standard quantum annealing scheme. 
%
%
Whether the identity of the exponents, both detected in quenches that emulate standard quantum annealing schemes, is merely coincidental or 
indicates a deeper link between the two phenomena, minimization of KZM defects and maximization of QFI and WYSI in the presence of noise, remains an open question.  
Moreover, in the presence of noise, the lack of an appropriate analytical solution corresponding to numerically challenging problems limits the ability to link the scaling exponent to the well-known critical exponents associated with equilibrium phase transitions.

\section*{Acknowledgment}
This work is based upon research funded by Iran National Science Foundation (INSF) under project No. 4024561


\appendix
\section{Reduced Density Matrix, Fermionic Correlation Functions, and Concurrence}
\label{appA}

To evaluate quantum correlations, we employ the two-spin reduced density matrix 
$\varrho_{\ell,m}(t)$ for spins located at sites $\ell$ and $m = \ell + r$, which takes the form \cite{Barouch1971a,Sadiek2010}
%
\be
\bl
\label{eq:APB1}
\varrho_{\ell,m}(t) =
\begin{pmatrix}
 \rho_{11} & 0 & 0 & \rho_{14} \\
 0 & \rho_{22} & \rho_{23} & 0 \\
 0 & \rho_{23}^{\ast} & \rho_{33} & 0 \\
 \rho_{14}^{\ast} & 0 & 0 & \rho_{44} 
\end{pmatrix},
\el
\ee
%
where the matrix elements are expressed in terms of one- and two-point correlation functions \cite{Barouch1971a,Sadiek2010}.
%
\be
\bl
\label{eq:APB2}
\rho_{11} &= \left\langle M^z_{\ell}\right\rangle+\left\langle s^{z}_{\ell} s^{z}_{m} \right\rangle+\tfrac{1}{4}; \quad
\rho_{22} = \rho_{33}=-\left\langle s^{z}_{\ell} s^{z}_{m} \right\rangle+\tfrac{1}{4},\\
\rho_{44} &= - \left\langle M^z_{\ell}\right\rangle+\left\langle s^{z}_{\ell} s^{z}_{m} \right\rangle+\tfrac{1}{4},\\
\rho_{23} &= \left\langle s^{x}_{\ell} s^{x}_{m} \right\rangle+\left\langle s^{y}_{\ell} s^{y}_{m} \right\rangle+i(\langle s^{x}_{\ell} s^{y}_{m}\rangle-\langle s^{y}_{\ell} s^{x}_{m}\rangle),\\
\rho_{14} &= \left\langle s^{x}_{\ell} s^{x}_{m} \right\rangle-\left\langle s^{y}_{\ell} s^{y}_{m} \right\rangle-i(\langle s^{x}_{\ell} s^{y}_{m}\rangle+\langle s^{y}_{\ell} s^{x}_{m}\rangle).
\el
\ee
%
Note that in a time-independent magnetic field $h_0(t) = h$, the two-point spin correlations 
$\langle s_{\ell}^x s_m^y \rangle$ and $\langle s_{\ell}^y s_m^x \rangle$ vanish. 
However, they are dynamically generated during a quench of the magnetic field. 
Using the Jordan--Wigner transformation, it has been shown that the two-point spin--spin correlation functions can be written as \cite{Barouch1971a,Sadiek2010}
%
\be
\bl
\label{eq:APB3}
S_{r}^{xx} = \langle s_1^x s_{1 + r}^x \rangle  
&= \tfrac{1}{4} \langle B_1 A_2 B_2 \ldots A_r B_r A_{r+1} \rangle, \\
S_{r}^{yy} = \langle s_1^y s_{1 + r}^y \rangle  
&= \tfrac{(-1)^n}{4} \langle A_1 B_2 A_2 \ldots B_r A_r B_{r+1} \rangle, \\
S_{r}^{zz} = \langle s_1^z s_{1 + r}^z \rangle  
&= \tfrac{1}{4} \langle A_1 B_1 A_{r+1} B_{r+1} \rangle, \\
S_{r}^{xy} = \langle s_1^x s_{1 + r}^y \rangle   
&= -\tfrac{i}{4} \langle B_1 A_2 B_2 \ldots A_r B_r B_{r+1} \rangle, \\
S_{r}^{yx} = \langle s_1^y s_{1 + r}^x \rangle   
&= \tfrac{i(-1)^r}{4} \langle A_1 B_2 A_2 \ldots B_r A_r A_{r+1} \rangle.
\el
\ee
%
where
\be
A_{\ell} = c_{\ell}^\dagger + c_{\ell}, 
\quad
B_{\ell} = c_{\ell}^\dagger - c_{\ell}.
\ee
%
It is straightforward to show that
%
\be
\bl
\label{eq:APB8}
\langle A_{\ell} A_m \rangle 
&= \langle c_{\ell}^\dagger c_m^\dagger \rangle 
  + \langle c_{\ell} c_m \rangle 
  + \langle c_{\ell}^\dagger c_m \rangle 
  + \langle c_{\ell} c_m^\dagger \rangle, \\
\langle B_{\ell} B_m \rangle 
&= \langle c_{\ell}^\dagger c_m^\dagger \rangle 
  + \langle c_{\ell} c_m \rangle 
  - \langle c_{\ell}^\dagger c_m \rangle 
  - \langle c_{\ell} c_m^\dagger \rangle, \\
\langle A_{\ell} B_m \rangle 
&= \langle c_{\ell}^\dagger c_m^\dagger \rangle 
  - \langle c_{\ell} c_m \rangle 
  - \langle c_{\ell}^\dagger c_m \rangle 
  + \langle c_{\ell} c_m^\dagger \rangle, \\
\langle B_{\ell} A_m \rangle
&= \langle c_{\ell}^\dagger c_m^\dagger \rangle 
  + \langle c_{\ell} c_m \rangle 
  - \langle c_{\ell}^\dagger c_m \rangle 
  - \langle c_{\ell} c_m^\dagger \rangle.
\el
\ee
%
Since the model is exactly solvable and can be written as a sum of $N/2$ non-interacting terms in momentum space, 
the four types of two-fermion correlation functions in real space [Eq.~(\ref{eq:APB8})] 
can be readily obtained by Fourier transforming to momentum space. For example,
%
\begin{equation} 
\label{Trace1}
\langle c_{\ell}^\dagger c_{\ell+r}^\dagger \rangle 
= \mathrm{Tr}(\rho\, c_{\ell}^\dagger c_{\ell+r}^\dagger)
= \frac{1}{N} \sum_{p,q} e^{-i(p\ell + q(\ell+r))} \langle c_p^\dagger c_q^\dagger \rangle.
\end{equation}

We use the diagonal basis $\{|\phi_k^\pm\rangle\}_k$ of the mode Hamiltonians at fixed time $t$ to evaluate the trace in Eq.~(\ref{Trace1}):  
%
\begin{equation} 
\label{eq:DiagTrace}
\mathrm{Tr} [\rho_kc_{k}^\dagger c_{-k}^\dagger]
=\langle\phi^{-}_k|\rho_kc_{k}^\dagger c_{-k}^\dagger|\phi^{-}_k\rangle + \langle\phi^{+}_k|\rho_kc_{k}^\dagger c_{-k}^\dagger|\phi^{+}_k\rangle.
\quad
\end{equation}
%
By inserting the resolution of identity 
$\mathbb{1} = |\phi^{-}_k\rangle\langle\phi^{-}_k| + |\phi^{+}_k\rangle\langle\phi^{+}_k|$ 
on the right-hand side, one obtains  
%
\begin{eqnarray} 
\label{eq:cc0}
\bl
\no
&\mathrm{Tr} 
[
\rho_k c_{k}^\dagger c_{-k}^\dagger
] 
=
\\
&
 \langle\phi^{-}_k|\rho_k|\phi^{-}_k\rangle
 \langle\phi^{-}_k|c_{k}^\dagger c_{-k}^\dagger|\phi^{-}_k\rangle
+ \langle\phi^{-}_k|\rho_k|\phi^{+}_k\rangle
\langle\phi^{+}_k|c_{k}^\dagger c_{-k}^\dagger|\phi^{-}_k\rangle 
 \\
&+ 
\langle\phi^{+}_k|\rho_k|\phi^{-}_k\rangle
\langle\phi^{-}_k|c_{k}^\dagger c_{-k}^\dagger|\phi^{+}_k\rangle
\!+\!
 \langle\phi^{+}_k|\rho_k|\phi^{+}_k\rangle
 \langle\phi^{+}_k|c_{k}^\dagger c_{-k}^\dagger|\phi^{+}_k\rangle 
  \\
 &= \rho^{(d)}_{k,11} ~\langle\phi^{-}_k|c_{k}^\dagger c_{-k}^\dagger|\phi^{-}_k\rangle 
 + 
\rho^{(d)}_{k,12} ~\langle\phi^{+}_k|c_{k}^\dagger c_{-k}^\dagger|\phi^{-}_k\rangle\\
&+
\rho^{(d)}_{k,21}  ~\langle\phi^{-}_k|c_{k}^\dagger c_{-k}^\dagger|\phi^{+}_k\rangle
+ \rho^{(d)}_{k,22}  ~\langle\phi^{+}_k|c_{k}^\dagger c_{-k}^\dagger|\phi^{+}_k\rangle, 
\el
\\
\end{eqnarray}  
%
where the superscript $(d)$ indicates that the density matrix is written in the diagonal basis. 
The matrix elements of $c_{k}^\dagger c_{-k}^\dagger$ can be computed from Eq.~(\ref{eq:APA5}), 
leading to
%
%
\begin{equation} 
\label{eq:cc1}
\bl
\langle c_{\l}^\dagger c_{\l+r}^\dagger \rangle = \frac{1}{N}\sum_{k>0} 
&
 \sin(kr) \Big[
\sin(2\theta_k) (\rho^{(d)}_{k,22} - \rho^{(d)}_{k,11}) 
\\&
+ 2i (\cos^2(\theta_k) \,\rho^{(d)}_{k,12} + \sin^2(\theta_k) \,\rho^{(d)}_{k,21} )
 \Big].
\el
\end{equation}
%
Similarly, one finds
%
%
\bea
\label{eq:cc2}
\bl
\langle c_{\l} c_{\l+r} \rangle =
 \frac{1}{N}\sum_{k>0} 
 &
\sin(kr)\Big[
\sin(2\theta_k) (\rho^{(d)}_{k,11} - \rho^{(d)}_{k,22}) 
\\
&
 +
 2i (\cos^2(\theta_k) 
\rho^{(d)}_{k,21} + \sin^2(\theta_k) \,\rho^{(d)}_{k,12} )
 \Big], 
 \no
 \el\\
\eea
\begin{equation} 
\bl
\label{eq:cc3}
\langle c_{\l}^\dagger c_{\l+r} \rangle 
=&
 \frac{1}{N}\sum_{k>0} 
\cos(kr) 
\Big[ 
2\cos^2(\theta_k) \rho^{(d)}_{k,22} + 2\sin^2(\theta_k) \rho^{(d)}_{k,11}
\\
& \quad\quad\quad
- i\sin(2\theta_k) (\rho^{(d)}_{k,12} - \rho^{(d)}_{k,21})
\Big], 
\el
\end{equation}
\begin{equation} 
\label{eq:cc4}
\bl
\langle c_{\l} c_{\l+r}^\dagger \rangle 
\!=\!
 \frac{1}{N}\sum_{k>0}
 &
   \cos(kr)
 \Big[
2\cos^2(\theta_k) \rho^{(d)}_{k,11} + 2\sin^2(\theta_k) \rho^{(d)}_{k,22}
  \\&
   - i\sin(2\theta_k) (\rho^{(d)}_{k,21} - \rho^{(d)}_{k,12})
    \Big].
   \el
\end{equation}
%
Substituting the results of Eqs.~(\ref{eq:cc1})--(\ref{eq:cc4}) into Eq.~(\ref{eq:APB8}), 
we obtain the $A$- and $B$-type two-point functions that determine the spin correlations 
in Eq.~(\ref{eq:amplitudes}). After simplification, the result reads
%
\begin{eqnarray} 
\label{eq:ABfinal}
\bl
\no
&\langle A_{l} A_{l+r} \rangle 
= \frac{2i}{N} \sum_{k>0} 
\mathrm{Re}[\rho^{(d)}_{k,12}]
\sin(kr) + \delta_{r,0}, 
\\
 &\langle B_{l} B_{l+r} \rangle 
 = \frac{2i}{N} \sum_{k>0} 
 \mathrm{Re}[\rho^{(d)}_{k,12}]
 \sin(kr) - \delta_{r,0},
  \\
 &\langle A_{l} B_{l+r} \rangle 
 = - \langle B_{l+r} A_{l} \rangle
 =
 \\
&
 \frac{1}{N} \sum_{k} \Big[
 (1-2\rho^{(d)}_{k,22})
\big[ \cos(kr)\cos(2\theta_{k})
\!-\!
 \sin(kr)\sin (2\theta_{k})
\big]
\\
&\quad\quad
- 2
\mathrm{Im}[\rho^{(d)}_{k,12}]
\big(\cos(kr)\sin(2\theta_{k}) + \sin(kr)\cos(2\theta_{k})\big) \Big].
\el
\\
\end{eqnarray}
%
We note that the two-point fermionic functions above reduce to their equilibrium forms 
if we set $\rho^{(d)}_{k,22} = \rho^{(d)}_{k,12} = \rho^{(d)}_{k,21} = 0$. In this case, we obtain
%
\bea
\bl
\langle A_{l}A_{l+r} \rangle =
& \delta_{r,0};
 \quad\quad 
\langle B_{l}B_{l+r} \rangle = -\delta_{r,0}, 
 \\
\langle A_{l} B_{l+r} \rangle =
&
 -\langle B_{l+r} A_{l} \rangle
\\  =&
 \frac{1}{N}\sum_{k}\Big[\cos(kr)\cos(2\theta_{k})-\sin(kr)\sin(2\theta_{k})\Big].
\no
\el
\\
\eea
%
In the thermodynamic limit $N \to \infty$, these correlation functions take the form
%
\be
\bl
&\langle {{A_l}{A_{l+r}}} \rangle = \delta(r);
 \quad\quad 
 \langle {{B_l}{B_{l+r}}} \rangle  -\delta(r), 
\\
&\langle A_{l} B_{l+r} \rangle = -\langle B_{l+r} A_{l} \rangle = \frac{1}{4\pi}\int_{0}^{2\pi} e^{ikr} e^{2i\theta_k}dk,
\quad\quad
\el
\ee
%
which is consistent with the results reported in Ref.~\cite{Franchini2017}.

Furthermore, for a quench from $h(t_{\text{initial}}) \to -\infty$ to $h(t_{\text{final}}) \to +\infty$, 
the off-diagonal elements of the density matrix, 
$\rho^{(d)}_{k,12}(t) = \big(\rho^{(d)}_{k,21}(t)\big)^{\ast}$, vanish, as does $\theta_{k}(t)$. 
Consequently, the two-point fermionic functions in Eq.~(\ref{eq:ABfinal}) reduce to
%
\be
\bl
\langle {{A_l}{A_{l+r}}} \rangle &= \delta_{r,0};
 \quad\quad 
\langle {{B_l}{B_{l+r}}} \rangle = -\delta_{r,0}, \\
\langle A_{l} B_{l+r} \rangle &= -\langle B_{l+r} A_{l} \rangle = \frac{1}{N}\sum_{k} (1- 2\rho^{(d)}_{k,22})\cos(kr),
\el
\ee
%
which, in the long-time thermodynamic limit, reproduces the results of Refs.~\cite{Sengupta2009,Nag2011,Cherng2006},
%
\be
\bl
\langle {{A_l}{A_{l+r}}} \rangle =& \delta(r);
 \quad\quad 
\langle {{B_l}{B_{l+r}}} \rangle = -\delta(r), \\
\langle A_{l} B_{l+r} \rangle=&-\langle B_{l+r} A_{l} \rangle 
= \frac{1}{2\pi} \int_{\pi}^{\pi} 
[1-2p_{k}(t)]
\cos(kr)dk\\
 = & -\frac{2}{\pi}\int_{0}^{\pi}p_{k}(t)\cos(kr)dk,
 \el
\ee
%
where 
\[
\lim_{t \to \infty} \rho^{(d)}_{k,22}(t) = e^{-4\pi \sin^{2}(k)\,\tau} \equiv p_{k}(t),
\]
with $\tau$ denoting the time scale of the quench 
\cite{Sengupta2009,Nag2011,Cherng2006}.


\twocolumngrid

\bibliography{QFI_References}

\end{document}